\documentclass{article}

\PassOptionsToPackage{numbers, compress}{natbib}

\usepackage[final]{neurips_2023}

\PassOptionsToPackage{hyphens}{url}

\usepackage[utf8]{inputenc} %
\usepackage[T1]{fontenc}    %
\usepackage{hyperref}       %
\usepackage{url}            %
\usepackage{booktabs}       %
\usepackage{amsfonts}       %
\usepackage{nicefrac}       %
\usepackage{microtype}      %
\usepackage[dvipsnames]{xcolor}

\usepackage{bm}
\usepackage[titlenumbered,linesnumbered,ruled]{algorithm2e}
\usepackage{xspace}
\usepackage{colortbl}
\usepackage{dcolumn}
\usepackage{mathtools}
\usepackage{tikz}  %
\usepackage{subcaption}
\usepackage{pgfplots}  %
\usepackage[squaren]{SIunits}
\usepackage{amsthm}
\usepackage{multirow}
\usepackage{natbib}
\usepackage{comment}
\usepackage{hhline}
\usepackage{amssymb}
\usepackage{nicematrix}

\newtheorem{theorem}{Theorem}
\newtheorem{lemma}[theorem]{Lemma}
\newtheorem{corollary}[theorem]{Corollary}
\newtheorem{proposition}[theorem]{Proposition}
\theoremstyle{definition}

\theoremstyle{remark}
\newtheorem{remark}[theorem]{Remark}

\newcommand\theDC{}
\newcommand\boldcell{\relax\ifmmode$\egroup\fi\bfseries\boldmath\theDC}
\makeatletter
\newcolumntype{E}[3]{>{\def\theDC{\DC@{#1}{#2}{#3}}\theDC}c<{\DC@end}}
\makeatother
\newcolumntype{d}[1]{E{.}{.}{#1}}
\newcolumntype{L}[1]{>{\raggedright\let\newline\\\arraybackslash\hspace{0pt}}m{#1}}
\newcolumntype{C}[1]{>{\centering\let\newline\\\arraybackslash\hspace{0pt}}m{#1}}
\newcommand\mc[1]{\multicolumn{1}{c}{#1}}
\newcommand{\bigcell}[2]{\begin{tabular}{@{}#1@{}}#2\end{tabular}}

\newenvironment{acks}{%
\section*{Acknowledgments}
}{%
}

\makeatletter
\if@submission
\excludecomment{acks}
\fi
\makeatother

\newcommand{\NA}{\ensuremath{\mathtt{NA}}}

\renewcommand{\vec}[1]{\bm{#1}}
\newcommand{\Xspace}{\ensuremath{\mathcal{X}}}
\newcommand{\Espace}{\ensuremath{\mathcal{E}}}

\newcommand{\Yspace}{\ensuremath{\mathcal{Y}}}
\newcommand{\Aspace}{\ensuremath{\mathcal{A}}}
\newcommand{\Fspace}{\ensuremath{\mathcal{F}}}

\newcommand{\del}{\ensuremath{\mathsf{del}}}
\newcommand{\ins}{\ensuremath{\mathsf{ins}}}
\newcommand{\sub}{\ensuremath{\mathsf{sub}}}
\newcommand{\ab}{\ensuremath{\mathsf{ab}}}
\newcommand{\edit}{\ensuremath{\epsilon}}
\newcommand{\pert}[1]{\bar{#1}}
\newcommand{\powerset}[1]{2^{#1}}
\newcommand{\radius}{r}
\newcommand{\smooth}[1]{#1}
\newcommand{\nbhd}{\mathcal{N}}
\newcommand{\base}[1]{{#1}_{\mathrm{b}}}
\newcommand{\reals}{\mathbb{R}}
\newcommand{\ind}[1]{\mathbf{1}_{#1}}
\newcommand{\umu}{\underline{\smash{\mu}}}
\newcommand{\citeteam}[1]{\citeauthor{#1}~\cite{#1}}
\makeatletter
\DeclareRobustCommand{\myvdots}{%
  \vbox{%
    \baselineskip 2.5\p@
    \lineskiplimit \z@
    \kern 3\p@
    \hbox{.}\hbox{.}\hbox{.}%
  }%
}
\makeatother

\newcommand{\ns}{$\mathsf{NS}$\xspace}
\newcommand{\rsdel}{$\mathsf{RS\text{-}Del}$\xspace}
\newcommand{\rsabn}{$\mathsf{RS\text{-}Abn}$\xspace}
\newcommand{\vtfeed}{VTFeed\xspace}
\newcommand{\sleipnir}{Sleipnir2\xspace}

\DeclareMathOperator{\LCS}{LCS}
\DeclareMathOperator{\dist}{dist}
\DeclareMathOperator{\apply}{apply}
\DeclareMathOperator{\ablate}{ablate}

\DeclareMathOperator*{\argmax}{arg\,max}

\DeclarePairedDelimiter\abs{\lvert}{\rvert}%
\DeclarePairedDelimiter\ceil{\lceil}{\rceil}
\DeclarePairedDelimiter\floor{\lfloor}{\rfloor}

\newcommand{\disp}{\textit{Disp}}
\newcommand{\slack}{\textit{Slack}}
\newcommand{\hdos}{\textit{HDOS}}
\newcommand{\hfield}{\textit{HField}}
\newcommand{\secinj}{\textit{GAMMA}}

\title{RS-Del: Edit Distance Robustness Certificates for Sequence Classifiers via Randomized Deletion}

\author{
  Zhuoqun Huang \\
  University of Melbourne \\
  \texttt{zhuoqun@unimelb.edu.au} \\
  \And
  Neil G.~Marchant \\
  University of Melbourne \\
  \texttt{nmarchant@unimelb.edu.au} \\
  \And
  Keane Lucas\\
  Carnegie Mellon University \\
  \texttt{keanelucas@cmu.edu} \\
  \AND
  Lujo Bauer \\
  Carnegie Mellon University \\
  \texttt{lbauer@cmu.edu} \\
  \And
  Olga Ohrimenko \\
  University of Melbourne \\
  \texttt{oohrimenko@unimelb.edu.au} \\
  \And 
  Benjamin I.~P.~Rubinstein \\
  University of Melbourne \\
  \texttt{brubinstein@unimelb.edu.au} \\
}

\begin{document}

\maketitle

\begin{abstract}
  Randomized smoothing is a leading approach for constructing classifiers that are certifiably robust against 
  adversarial examples. 
  Existing work on randomized smoothing has focused on classifiers with continuous 
  inputs, such as images, where $\ell_p$-norm bounded adversaries are commonly studied. 
  However, there has been limited work for classifiers with discrete or variable-size inputs,
  such as for source code, 
  which require different threat models and smoothing 
  mechanisms. 
  In this work, we adapt randomized smoothing for discrete sequence classifiers to provide certified robustness 
  against edit distance-bounded adversaries. 
  Our proposed smoothing mechanism \emph{randomized deletion} (\rsdel) applies random deletion edits, which are (perhaps surprisingly) sufficient to 
  confer robustness against adversarial deletion, insertion and substitution edits. 
  Our proof of certification deviates from the established Neyman-Pearson approach, which is intractable in our setting, and is instead organized around longest common subsequences. 
  We present a case study on malware detection---a binary classification problem on byte sequences where classifier evasion is a well-established threat model.  
  When applied to the popular MalConv malware detection model, our smoothing mechanism \rsdel\ achieves a certified 
  accuracy of 91\% at an edit distance radius of 128~bytes. 
\end{abstract}

\section{Introduction}
\label{sec:intro}

Neural networks have achieved exceptional performance for classification tasks in unstructured domains, such 
as computer vision and natural language. 
However, they are susceptible to adversarial examples---inconspicuous perturbations that cause inputs to be 
misclassified~\cite{szegedy2014intriguing,goodfellow2015explaining}. 
While a multitude of defenses have been proposed against adversarial examples, they have historically been broken by 
stronger attacks.  
For instance, six of nine defense papers accepted for presentation at ICLR2018 were defeated months before the 
conference took place~\cite{athalye2018obfuscated}; another tranche of thirteen defenses were circumvented shortly 
after~\cite{tramer2020adaptive}. 
This arms race between attackers and defenders has led the community to investigate certified robustness, 
which aims to guarantee that a classifier's prediction cannot be changed under a specified set of adversarial 
perturbations~\cite{wong2018provable,raghunathan2018certified}. 

Most prior work on certified robustness has focused on classifiers with \emph{continuous fixed-dimensional} inputs, 
under the assumption of $\ell_p$-norm bounded perturbations. 
A variety of certification methods have been proposed for this setting, including deterministic methods which 
bound a network's output by convex relaxation~\cite{wong2018provable,dvijotham2018training,
raghunathan2018certified,weng2018towards} or composing layerwise bounds~\cite{mirman2018differentiable,weng2018towards,
tsuzuku2018lipschitz,gowal2019effectiveness,zhang2020towards}, and randomized smoothing which obtains 
high probability bounds for a base classifier smoothed under random noise~\cite{lecuyer2019certified,cohen2019certified,
yang2020randomized}. 
Randomized smoothing has achieved state-of-the-art certified robustness against $\ell_2$-norm bounded perturbations 
on ImageNet, correctly classifying 71\% of a test set under perturbations with $\ell_2$-norm up to $127/255$ 
(half maximum pixel intensity)~\cite{carlini2023certified}. 
However, real-world attacks go beyond continuous modalities: a range of attacks have been demonstrated against models with 
discrete inputs, such as binary executables~\cite{lucas2021malware,kolosnjaji2018adversarial,park2019generation,kreuk2019deceiving},
source code~\cite{zhang2020generating}, and PDF files~\cite{chen2020training}.

Unfortunately, there has been limited investigation of certified robustness for discrete modalities, where prior work has focused 
on perturbations with bounded Hamming ($\ell_0$) distance~\cite{lee2019tight,levine2020robustness,jia2022almost}, in some cases 
under additional constraints~\cite{jia2019certified,ye2020safer,ren2019generating,saha2023adversarial,moon2023randomized}. 
While this work covers attackers that overwrite content, it does not cover attackers that insert\slash delete content, 
which is important for instance in the malware domain~\cite{lucas2021malware,demetrio2021functionality}.
One exception is \citeauthor{zhang2021certified}'s method for certifying robustness under string transformation 
rules that may include insertion and deletion, however their approach is only feasible for small rule sets and 
is limited to LSTMs~\cite{zhang2021certified}. 
\citeteam{liu2021pointguard} also study certification for a threat model that includes insertion\slash 
deletion in the context of point cloud classification.

In this paper we develop a comprehensive treatment of \emph{edit distance certifications} for sequence classifiers. 
We consider input sequences of varying length on finite domains. 
We cover threat models where an adversary can arbitrarily perturb sequences by substitutions, insertions, and deletions or any subset of these operations. Moreover,
our framework encompasses adversaries that apply edits to blocks of tokens at a time. 
Such threat models are motivated in malware analysis, where attackers are likely to edit semantic objects in an executable (e.g., whole instructions), rather than at the byte-level. 
We introduce tunable decision thresholds, %
to adjust decision boundaries while still forming sound certifications. This permits trading-off misclassification rates and certification radii between classes, and is useful in settings 
where adversaries have a strong incentive to misclassify malicious instances as benign~\cite{romeo2018adversarially,pfrommer2023asymmetric}. 

We accomplish our certifications using randomized smoothing~\cite{lecuyer2019certified,cohen2019certified}, which we instantiate with a general-purpose deletion smoothing mechanism called \rsdel. 
Perhaps surprisingly, \rsdel does not need to sample all edit operations covered by our certifications. 
By smoothing using deletions only, we simplify our certification analysis and achieve the added benefit of improved efficiency by querying the base classifier with shorter sequences. 
\rsdel is compatible with arbitrary base classifiers, requiring only oracle query access such as via an inference API. 
To prove our robustness certificates, we have to deviate from the standard Neyman-Pearson approach to randomized smoothing, as it is intractable in our setting. 
Instead, we organize our proof around a representative longest common subsequence (LCS):
the LCS serves as a reference point in the smoothed space between an input instance and a neighboring instance, allowing us to bound the confidence of the smoothed model between the two. 

Finally, we present a comprehensive evaluation of \rsdel in the malware analysis setting.\footnote{Our implementation is available at \url{https://github.com/Dovermore/randomized-deletion}.} 
We investigate tradeoffs between accuracy and the size of our certificates for varying levels of deletion, and  
observe that \rsdel can achieve a certified accuracy of 91\% at an edit distance radius of 128~bytes 
 using on the order of $10^3$ model queries. By comparison, a brute force certification would require in excess of $10^{308}$ model queries to certify the same radius.  
We also demonstrate asymmetric certificates (favouring the malicious class) and certificates covering edits at the 
machine instruction level by leveraging chunking and information from a disassembler. 
Finally, we assess the empirical robustness of \rsdel\ to several attacks 
from the literature, where we observe a reduction in attack success rates.

\section{Preliminaries} \label{sec:formulation}

\paragraph{Sequence classification}
Let $\Xspace=\Omega^\star$ represent the space of finite-length sequences (including the empty sequence) whose elements are drawn 
from a finite set $\Omega$. 
For a sequence $\vec{x} \in \Xspace$, we denote its length by $\abs{\vec{x}}$ and the element at position $i$ by 
$x_i$ where $i$ runs from $1$ to $\abs{\vec{x}}$. 
We consider classifiers that map sequences in $\Xspace$ to $K$ classes in $\Yspace=\{0,\ldots,K-1\}$. 
For example, in our case study on malware detection, we take $\Xspace$ to be the space of byte sequences (binaries), and
$\Omega$ to be the set of bytes and $\Yspace$ to be $\{0, 1\}$ where 0 and 1 denote benign and malicious binaries 
respectively.

\paragraph{Robustness certification}

Given a classifier $f \colon \Xspace \to \Yspace$, an input $\vec{x} \in \Xspace$ and a neighborhood 
$\nbhd(\vec{x}) \subset \Xspace$ around $\vec{x}$, a \emph{robustness certificate} is a guarantee that the classifier's 
prediction is constant in the neighborhood, i.e., 
\begin{equation}
  f(\vec{x}) = f(\vec{x}') \quad \forall \vec{x}' \in \nbhd(\vec{x}).  \label{eqn:robustness-cert}
\end{equation}
When the neighborhood corresponds to a ball of radius $\radius$ centered on $\vec{x}$, we make the dependence on 
$\radius$ explicit 
by writing $\nbhd_r(\vec{x})$. 
We adopt the paradigm of \emph{conservative}, \emph{probabilistic} certification, which is a natural fit for randomized 
smoothing~\cite{cohen2019certified}. 
Under this paradigm, a certifier may either assert that \eqref{eqn:robustness-cert} holds with high probability, 
or decline to assert whether \eqref{eqn:robustness-cert} holds or not.

\paragraph{Edit distance robustness}

Most existing work on robustness certification focuses on classifiers that operate on \emph{fixed-dimensional} inputs 
in $\reals^d$, where the neighborhood of certification is an $\ell_p$-ball \cite{raghunathan2018certified,
wong2018provable,lecuyer2019certified,cohen2019certified,yang2020randomized}. 
However, $\ell_p$ robustness is not well motivated for sequence classifiers: 
$\ell_p$ neighborhoods are limited to constant-length sequences, and the norm is ill-defined for 
sequences with non-numeric elements. 
For example, an $\ell_p$ neighborhood around a byte sequence like 
$\vec{x} = (78, 7\mathrm{A}, 2\mathrm{D}, 00)$ must include sequences additively perturbed by real-valued sequences 
like $\vec{\delta} = (-0.09,  0.07, 0.01,  0.1)$, which clearly results in a type mismatch.
Even if one focuses on robustness for length-preserving sequence perturbations, the $\ell_p$ neighborhood is a poor 
choice because it excludes sequences that are even slightly misaligned. 
Motivated by these shortcomings, we consider \emph{edit distance} robustness. 

Edit distance is a natural measure for comparing sequences. 
Given a set of elementary edit operations (ops) $O$, the edit distance $\dist_O(\vec{x}, \vec{x}')$ is the minimum 
number of ops required to transform sequence $\vec{x}$ into $\vec{x}'$. 
We consider three ops: delete a single element (\del), insert a single element (\ins), and substitute one element with 
another (\sub). 
For generality, we allow $O$ to be any combination of these ops.
When $O = \{\del, \ins, \sub\}$ the edit distance is known as Levenshtein distance~\cite{levenshtein1966binary}. 
A primary goal of this paper is to produce edit distance robustness certificates, where the neighborhood of certification 
is an edit distance ball $\nbhd_r(\vec{x}) = \{\vec{x}' \in \Xspace: \dist_O(\vec{x}', \vec{x}) \leq \radius\}$.

\paragraph{Threat model}
We consider an adversary that has full knowledge of our base and smoothed models, source of randomness,  certification scheme, and possesses unbounded computation. The attacker makes edits from $O$ up to some budget, in order to misclassify a target $\vec{x}$.
In the context of our experimental case study on malware detection, edit distance is a reasonable proxy for the 
cost of running evasion attacks that iteratively apply localized functionality-preserving edits (e.g., 
\cite{park2019generation,demetrio2019explaining,nisi2021lost,lucas2021malware,song2022mab}). 
Since the edit distance scales roughly linearly with the number of attack iterations, the  
adversary has an incentive to minimize edit distance for these attacks.

\section{RS-Del: Randomized deletion smoothing} \label{sec:method}

In this section, we propose a method for constructing sequence classifiers that are certifiably robust under 
bounded edit distance perturbations. Our method \rsdel extends randomized smoothing with a novel deletion mechanism and 
tunable decision thresholds.
We review randomized smoothing in Section~\ref{sec:rs} and describe our deletion mechanism in Section~\ref{sec:del-mech}, along
with its practical aspects in Section~\ref{sec:practicalities}.
We summarize the certified robustness guarantees of our method in Table~\ref{tbl:certificate-O}
and defer their derivation to Section~\ref{sec:certificate}.

\subsection{Randomized smoothing} \label{sec:rs}

Let $\base{f}: \Xspace \to \Yspace$ be a \emph{base} classifier and $\phi \colon \Xspace \to \mathcal{D}(\Xspace)$ 
be a \emph{smoothing mechanism} that maps inputs to a distributions over (perturbed) inputs. 
Randomized smoothing composes $\base{f}$ and $\phi$ to construct a new \emph{smoothed} classifier 
$\smooth{f} \colon \Xspace \to \Yspace$. 
For any input $\vec{x} \in \Xspace$, the smoothed classifier's prediction is
\begin{equation}
    \smooth{f}(\vec{x}) := \argmax_{y \in \Yspace} \left\{ p_y(\vec{x}; \base{f}, \phi) - \eta_y \right\},  \label{eqn:smoothed-classifier}
\end{equation}
where $\vec{\eta} = \{\eta_y\}_{y \in \Yspace}$ is a set of real-valued tunable decision thresholds and 
\begin{equation}
    p_y(\vec{x}; \base{f}, \phi) = \Pr_{\vec{z} \sim \phi(\vec{x})} \left[\base{f}(\vec{z}) = y\right] \label{eqn:smoothed-prob}
\end{equation}
is the probability that the base classifier $\base{f}$ predicts class $y$ for a perturbed input drawn from 
$\phi(\vec{x})$. 
We omit the dependence of $p_y$ on $\base{f}$ and $\phi$ where it is clear from context.

The viability of randomized smoothing as a method for achieving certified robustness is strongly dependent on the 
smoothing mechanism. 
Ideally, the mechanism should be chosen to yield a smoothed classifier with improved robustness under the chosen threat 
model, while minimizing any drop in accuracy compared to the base classifier. 
The mechanism should also be amenable to analysis, so that a tractable robustness certificate can be derived. 

\begin{remark}
  Previous definitions of randomized smoothing (e.g.,~\cite{lecuyer2019certified,cohen2019certified}) do not 
  incorporate decision thresholds, and effectively assume $\vec{\eta} = 0$. 
  We introduce decision thresholds as a way to trade off error rates and robustness between classes. 
  This is useful when there is asymmetry in misclassification costs across classes. 
  For instance, in our case study on malware detection, robustness of benign examples is less important because 
  adversaries have limited incentive to trigger misclassification of benign examples~\cite{romeo2018adversarially,
  pfrommer2023asymmetric,fleshman2019nonnegative}.
  We note that the base classifier may also be equipped with decision thresholds, which provide another degree of 
  freedom to trade off error rates between classes. 
\end{remark}

\subsection{Randomized deletion mechanism} \label{sec:del-mech}

We propose a smoothing mechanism that achieves certified edit distance robustness. 
Our smoothing mechanism perturbs a sequence by deleting elements at random, and is called \emph{randomized deletion}, 
or \emph{\rsdel} for short. 

Consider a sequence $\vec{x} \in \Xspace$ whose elements are indexed by the set 
$[\vec{x}] = \{1, \ldots, \abs{\vec{x}}\}$. 
We specify the distribution of $\phi(\vec{x})$ for \rsdel\ in two stages. 
In the first stage, a random edit $\edit$ is drawn from a distribution $G(\vec{x})$ over the space of possible edits 
to $\vec{x}$, denoted $\Espace(\vec{x})$. 
Since we only consider deletions for smoothing, any edit can be represented by the set of element indices in $[\vec{x}]$ that 
\emph{remain} after deletion. 
Hence $\Espace(\vec{x})$ is taken as the powerset of $[\vec{x}]$. 
We specify edit distribution $G(\vec{x})$ so that each element is deleted i.i.d.\ with probability $p_\del \in (0, 1)$:%
\begin{equation}
    \Pr[G(\vec{x}) = \edit] = \prod_{i = 1}^{\abs{\vec{x}}} p_\del^{\ind{i \notin \edit}} (1 - p_\del)^{\ind{i \in \edit}},
    \label{eqn:edit-distribution}
\end{equation}
where $\ind{A}$ denotes the indicator function, which returns 1 if $A$ evaluates to true and 0 otherwise. 
In the second stage, the edit $\edit$ is applied to $\vec{x}$ to yield the perturbed sequence:
\begin{equation}
    \vec{z} = \apply(\vec{x}, \edit) \coloneqq \left( x_{\edit_{(i)}} \right)_{i = 1 \ldots \abs{\edit}},
    \label{eqn:apply-edit}
\end{equation} 
where $\edit_{(i)}$ denotes the $i$-th smallest index in $\edit$.
The perturbed sequence $\vec{z}$ is guaranteed to be a subsequence of $\vec{x}$. 
Putting both stages together, the distribution of $\phi(\vec{x})$ is
\begin{equation}
    \Pr[\phi(\vec{x}) = \vec{z}] 
    = \sum_{\edit \in \Espace(\vec{x})} \Pr[G(\vec{x}) = \edit] \ind{\apply(\vec{x}, \edit) = \vec{z}}.
    \label{eqn:del-mechanism}
\end{equation}

\begin{remark}
    It may be surprising that we are proposing a smoothing mechanism for certified edit distance robustness that 
    does not use the full set of edit ops $O$ covered by the threat model. 
    It is a misconception that randomized smoothing requires perfect alignment between the mechanism and the threat 
    model. 
    All that is needed from a robustness perspective, is for the mechanism to return distributions that are 
    statistically close for any pair of inputs that are close in $O$ edit distance; this can be achieved solely with 
    deletion. 
    In fact, perfect alignment is known to be suboptimal for some~$\ell_p$ threat models~\cite{yang2020randomized}. 
    Our deletion mechanism leads to a tractable robustness certificate covering the full set of edit ops (see 
    Section~\ref{sec:certificate}). 
    Moreover while benefiting robustness, our empirical results show that our deletion mechanism has only a minor 
    impact on accuracy (see Section~\ref{sec:evaluation}). 
    Finally, our deletion mechanism reduces the length of the input, which is beneficial for computational 
    efficiency (see Appendix~\ref{app-sec:eval-comp-efficiency}). 
    This is not true in general for mechanisms employing insertions\slash substitutions. 
\end{remark}

\subsection{Practical considerations} \label{sec:practicalities}

We now discuss considerations for implementing and certifying \rsdel\ in practice. 
In doing so, we reference theoretical results for certification, which are covered later in
Section~\ref{sec:certificate}.

\paragraph{Probabilistic certification} 
Randomized smoothing does not generally support exact evaluation of the classifier's confidence 
$\mu_y := p_y(\vec{x})$, which is required for exact prediction and exact evaluation of the certificates we develop 
in Section~\ref{sec:certificate}. 
While $\mu_y$ can be evaluated exactly for \rsdel by enumerating over the possible edits $\Espace(\vec{x})$, 
the computation scales exponentially in $\abs{\vec{x}}$ (see Appendix~\ref{app-sec:exact}). 
Since this is infeasible for even moderately-sized $\vec{x}$, we follow standard practice in randomized smoothing 
and estimate $\mu_y$ with a lower confidence bound using Monte Carlo sampling~\cite{cohen2019certified}. 
This procedure is described in pseudocode in Figure~\ref{alg:estimation-algorithm}: lines~1--3 estimate the 
predicted class $y = \smooth{f}(\vec{x})$, lines~4--5 compute a $1 - \alpha$ lower confidence bound on $\mu_y$, and
line~6 uses this information and the results in Table~\ref{tbl:certificate-O} to compute a probabilistic certificate 
that holds with probability $1 - \alpha$.
If the lower confidence bound on $\mu_y$ exceeds the corresponding detection threshold $\eta_y$, the prediction and 
certificate are returned (line~8), otherwise we \emph{abstain} due to lack of statistical significance (line~7).

\paragraph{Training} 
While randomized smoothing is compatible with any base classifier, it generally performs poorly for 
conventionally-trained classifiers~\cite{lecuyer2019certified}. 
We therefore train base classifiers specifically to be used with \rsdel, by replacing original sequences with perturbed 
sequences (drawn from $\phi$) at training time. 
This has been shown to achieve good empirical performance in prior work~\cite{cohen2019certified}. 

\paragraph{Sequence chunking} 
So far, we have described edit distance robustness and \rsdel\ assuming edits are applied at the level of sequence 
elements. 
However in some applications it may be reasonable to assume edits are applied at a coarser level, to contiguous 
chunks of sequence elements. 
For example, in malware analysis, one can leverage information from a disassembler to group low-level sequence 
elements (bytes) into more semantically meaningful chunks, such as machine instructions, addresses and header 
fields (see Appendix~\ref{sec:malware-detection}). 
Our methods are compatible with chunking---we simply reinterpret the sequence as a sequence of chunks, rather than 
a sequence of lower-level elements. 
This can yield a tighter robustness certificate, since edits within chunks are excluded.

\begin{figure}
  \begin{minipage}[t]{.535\textwidth}
    \vspace{0pt}
    \IncMargin{1em}
    \setlength{\interspacetitleruled}{0pt}%
    \setlength{\algotitleheightrule}{0pt}%
    \begin{algorithm}[H]
    \DontPrintSemicolon
    Sample perturbed inputs for prediction estimate: $S \gets \{\vec{z}_i \sim \phi(\vec{x})\}_{i = 1 \ldots n_\mathrm{pred}}$\;
    Confidence scores: $\hat{\mu}_y \gets \frac{1}{\abs{S}} \sum_{\vec{z} \in S} \ind{\base{f}(\vec{z}) = y}$\;
    Prediction: $\hat{y} \gets \argmax_{y \in \Yspace} \{\hat{\mu}_y - \eta_y\}$\;
    Sample perturbed inputs for lower confidence bound (LCB): $S' \gets \{\vec{z}_i \sim \phi(\vec{x})\}_{i = 1 \ldots n_\mathrm{bnd}}$\;
    LCB: $\umu_{\hat{y}} \gets \texttt{BinLCB}(\sum_{\vec{z} \in S'}\ind{\base{f}(\vec{z}) = \hat{y}}, \abs{S'}, \alpha)$\;
    Compute certified radius $\radius^\star$ using Table~\ref{tbl:certificate-O}\;
    \lIf{$\umu_{\hat{y}} < \eta_{\hat{y}}$}{
      \Return{abstain}
    }
    \lElse{
      \Return{prediction~$\hat{y}$, radius~$\radius^\star$}
    }
    \end{algorithm}
    \DecMargin{1em}
    \captionof{figure}{Probabilistic certification of \rsdel. 
    Here $\vec{x}$ is the input sequence, $\base{f}$ is the base classifier, $p_\del$ is the deletion probability, 
    $\vec{\eta}$ is the set of decision thresholds, $\alpha$ is the significance level, and $n_\mathrm{pred}, 
    n_\mathrm{bnd}$ are sample sizes. 
    $\texttt{BinLCB}(k,n,\alpha)$ returns a lower confidence bound for $p$ at level $\alpha$ given 
    $k \sim \operatorname*{Bin}(n, p)$.}
    \label{alg:estimation-algorithm}
  \end{minipage}
  \hfill
  \begin{minipage}[t]{0.44\textwidth}
    \vspace{3pt}
    \centering
    \begin{tabular}{cccc}
      \toprule
      \multicolumn{3}{c}{Edit ops $O$}     %
          &  \\
      \cmidrule(lr){1-3}
      \del       & \ins       & \sub       %
          & Certified radius $\radius^\star$ \\
      \midrule
                 & \checkmark &            %
          & $\floor*{ \vcenter{\hbox{$\nicefrac{\log \frac{1 - \mu_y}{1 - \nu_y}}{\log p_\del}$}} }$ \\[2pt]
      \checkmark &            &            %
          & $\floor*{ \nicefrac{\log \frac{\nu_y}{\mu_y}}{\log p_\del} }$ \\[2pt]
      \checkmark & \checkmark &            %
          & $\floor*{ \nicefrac{\log \frac{\nu_y}{\mu_y}}{\log p_\del} }$ \\[2pt]
      \checkmark & \checkmark & \checkmark %
          & $\floor*{ \nicefrac{\log (1 + \nu_y - \mu_y)}{\log p_\del} }$ \\[2pt]
                 &            & \checkmark %
          & $\floor*{ \nicefrac{\log (1 + \nu_y - \mu_y)}{\log p_\del} }$ \\[2pt]
                 & \checkmark & \checkmark %
          & $\floor*{ \nicefrac{\log (1 + \nu_y - \mu_y)}{\log p_\del} }$ \\[2pt]
      \checkmark &            & \checkmark %
          & $\floor*{ \nicefrac{\log (1 + \nu_y - \mu_y)}{\log p_\del} }$ \\[2pt]
      \bottomrule  
    \end{tabular}
    \captionof{table}{Edit distance robustness certificates for \rsdel\ as a function of the edit ops $O$ used to 
      define the edit distance. Here $\mu_y$ is the confidence for predicted class $y$ and $\nu_y$ is a 
      threshold derived from $\vec{\eta}$ defined in \eqref{eqn:cert-conf-thresh}.}
    \label{tbl:certificate-O}
    \end{minipage}
\end{figure}

\section{Edit distance robustness certificate} \label{sec:certificate}

We now derive an edit distance robustness certificate for \rsdel. 
We present the derivation in three parts: Section~\ref{sec:cert-outline} provides an outline, 
Section~\ref{sec:lb-prob-score} derives a lower bound on \rsdel's confidence score
and Section~\ref{sec:regional} uses the bound to complete the derivation.
All proofs are presented in Appendix~\ref{app-sec:proofs}.

\subsection{Derivation outline} \label{sec:cert-outline}

Following prior work~\cite{cohen2019certified,lecuyer2019certified,dvijotham2020framework}, we derive an edit distance 
robustness certificate that relies on limited information about \rsdel. 
We allow the certificate to depend on 
the input $\vec{x}$, 
the smoothed prediction $y = \smooth{f}(\vec{x})$, 
the confidence score $\mu_y = p_y(\vec{x}; \base{f})$, 
the decision threshold $\eta_y$, and 
the architecture of $\smooth{f}$, including the deletion smoothing mechanism $\phi$, but excluding the architecture of 
the base classifier $\base{f}$.
Limiting the dependence in this way improves tractability and ensures that the certificate is applicable for any 
choice of base classifier $\base{f}$. 
Formally, the only information we assume about $\base{f}$ is that it is some classifier in the feasible base classifier set:
\begin{equation}
  \Fspace(\vec{x}, \mu_y) = \left\{\, h \in \Xspace \to \Yspace : \mu_y = p_y(\vec{x}; h) \,\right\}.
  \label{eqn:consistent-detectors}
\end{equation}

Recall that an edit distance robustness certificate at radius $\radius$ for a classifier $\smooth{f}$ at input 
$\vec{x}$ is a guarantee that $\smooth{f}(\vec{x}) = \smooth{f}(\pert{\vec{x}})$ for any perturbed input 
$\pert{\vec{x}}$ in the neighborhood 
$\nbhd_\radius(\vec{x}) = \{\pert{\vec{x}} \in \Xspace: \dist_O(\pert{\vec{x}}, \vec{x}) \leq \radius \}$. 
We observe that this guarantee holds for \rsdel\ in the limited information setting iff the $\vec{\eta}$-adjusted 
confidence for predicted class $y$ exceeds the $\vec{\eta}$-adjusted confidence for any other class for all 
perturbed inputs $\pert{\vec{x}}$ and feasible base classifiers $h$:
\begin{equation}
  p_y(\pert{\vec{x}}; h) - \eta_{y} \geq \max_{y' \neq y} \{ p_{y'}(\pert{\vec{x}}; h) - \eta_{y'} \} \quad 
    \forall \pert{\vec{x}} \in \nbhd_\radius(\vec{x}), h \in \Fspace(\vec{x}, \mu_y).
  \label{eqn:rsdel-cert}
\end{equation}
To avoid dependence on the confidence of the runner-up class, which is inefficient for probabilistic certification, 
we work with the following more convenient condition.
\begin{proposition} \label{prop:cert-suff}
  A sufficient condition for \eqref{eqn:rsdel-cert} is $\rho(\vec{x}, \mu_y) \geq \nu_y(\vec{\eta})$ where 
  \begin{equation}
    \rho(\vec{x}, \mu_y) \coloneqq \min_{\pert{\vec{x}} \in \nbhd_\radius(\vec{x})} 
        \min_{h \in \Fspace(\vec{x}, \mu_y)} p_y(\pert{\vec{x}}; h)
    \label{eqn:regional-exact}
  \end{equation}
  is a tight lower bound on the confidence for class $y$, and we define the threshold
  \begin{equation}
    \nu_y(\vec{\eta}) = \begin{cases}
      \frac{1}{2} + \eta_y - \min_{y' \neq y} \eta_{y'}, 
        & \eta_y \geq \min_{y' \neq y} \eta_{y'} \text{ and } \abs{\Yspace} > 2, \\
      1 + \eta_y - \min_{y' \neq y} \eta_{y'}, 
        & \eta_y < \min_{y' \neq y} \eta_{y'} \text{ and } \abs{\Yspace} > 2, \\
      \frac{1 + \eta_y - \min_{y' \neq y} \eta_{y'}}{2}, 
        & \abs{\Yspace} = 2.
    \end{cases}
    \label{eqn:cert-conf-thresh}
  \end{equation}
\end{proposition}

The standard approach for evaluating $\rho(\vec{x}, \mu_y)$ is via the Neyman-Pearson 
lemma~\cite{neyman1933,cohen2019certified}, however this seems insurmountable in our setting due to the challenging 
geometry of the edit distance neighborhood. 
We therefore proceed by deriving a loose lower bound on the confidence 
$\tilde{\rho}(\vec{x}, \mu_y) \leq \rho(\vec{x}, \mu_y)$, noting that the robustness guarantee still holds so long as 
$\tilde{\rho}(\vec{x}, \mu_y) > \nu_y(\vec{\eta})$.
The derivation proceeds in two steps. 
In the first step, covered in Section~\ref{sec:lb-prob-score}, we derive a lower bound for the inner minimization 
in \eqref{eqn:regional-exact}, which we denote by $\tilde{\rho}(\pert{\vec{x}}, \vec{x}, \mu_y)$.  
Then in the second step, covered in Section~\ref{sec:regional}, we complete the derivation by minimizing 
$\tilde{\rho}(\pert{\vec{x}}, \vec{x}, \mu_y)$ over the edit distance neighborhood. 
Our results are summarized in Table~\ref{tbl:certificate-O}, where we provide certificates under various constraints 
on the edit ops.

\subsection{Minimizing over feasible base classifiers} \label{sec:lb-prob-score}

In this section, we derive a loose lower bound on the classifier confidence with respect to feasible base 
classifiers
\begin{equation}
  \tilde{\rho}(\pert{\vec{x}}, \vec{x}, \mu_y) \leq \rho(\pert{\vec{x}}, \vec{x}, \mu_y) = \min_{h \in \Fspace(\vec{x}, \mu_y)} p_y(\pert{\vec{x}}; h).
\end{equation}
To begin, we write $p_y(\pert{\vec{x}}; h)$ as a sum over the edit space by combining \eqref{eqn:smoothed-prob}
and \eqref{eqn:del-mechanism}:
\begin{equation}
    p_y(\pert{\vec{x}}; h) = \sum_{\pert{\edit} \in \Espace(\pert{\vec{x}})} s(\pert{\edit}, \pert{\vec{x}}; h) 
    \quad \text{where} \quad s(\pert{\edit}, \pert{\vec{x}}; h) = \Pr\left[G(\pert{\vec{x}}) = \pert{\edit}\right] 
        \ind{h(\apply(\pert{\vec{x}}, \pert{\edit})) = y}. \label{eqn:smoothed-prob-summand}
\end{equation}
We would like to rewrite this sum in terms of the known confidence score at $\vec{x}$, 
$\mu_y = p_y(\vec{x}; h) 
= \sum_{\edit \in \Espace(\vec{x})} s(\edit, \vec{x}; h)$. 
To do so, we identify pairs of edits $\pert{\edit}$ to $\pert{\vec{x}}$ and $\edit$ to $\vec{x}$ for which the 
corresponding terms $s(\pert{\edit}, \pert{\vec{x}}; h)$ and $s(\edit, \vec{x}; h)$ are proportional. 

\begin{lemma}[Equivalent edits] \label{lem:equiv-edits}
    Let $\vec{z}^\star$ be a longest common subsequence (LCS) \cite{wagner1974string}
     of $\pert{\vec{x}}$ and $\vec{x}$, and let 
    $\pert{\edit}^\star \in \Espace(\pert{\vec{x}})$ and $\edit^\star \in \Espace(\vec{x})$ be any edits such that 
    $\apply(\pert{\vec{x}}, \pert{\edit}^\star) = \apply(\vec{x}, \edit^\star) = \vec{z}^\star$. 
    Then there exists a bijection $m: \powerset{\pert{\edit}^\star} \to \powerset{\edit^\star}$ such that 
    $\apply(\pert{\vec{x}}, \pert{\edit}) = \apply(\vec{x}, \edit)$ for any $\pert{\edit} \subseteq \pert{\edit}^\star$ and 
    $\edit = m(\pert{\edit})$.
    Furthermore, we have $s(\pert{\edit}, \pert{\vec{x}}; h) = 
    p_\del^{\abs{\pert{\vec{x}}} - \abs{\vec{x}}} s(\edit, \vec{x}; h)$.
\end{lemma}

Applying this proportionality result to all pairs of edits $\pert{\edit}, \edit$ related under the bijection $m$ 
yields: 
\begin{equation*}
    \sum_{\pert{\edit} \in \powerset{\pert{\edit}^\star}} s(\pert{\edit}, \pert{\vec{x}}; h) 
        = p_\del^{\abs{\pert{\vec{x}}} - \abs{\vec{x}}} 
            \sum_{\edit \in \powerset{\edit^\star}} s(\edit, \vec{x}; h).
\end{equation*}
Thus we can achieve our goal of writing $p_y(\pert{\vec{x}}; h)$ in terms of $\mu_y$. 
A rearrangement of terms gives:
\begin{equation}
    \begin{split}
    p_y(\pert{\vec{x}}; h) &=  p_\del^{\abs{\pert{\vec{x}}} - \abs{\vec{x}}} \left(
            \mu_y - \sum_{\edit \notin \powerset{\edit^\star}} s(\edit, \vec{x}; h) 
        \right) %
          + \sum_{\pert{\edit} \notin \powerset{\pert{\edit}^\star}} s(\pert{\edit}, \pert{\vec{x}}; h).
    \end{split}
    \label{eqn:smooth-prob-mu}
\end{equation}
This representation is convenient for deriving a lower bound. 
Specifically, we can drop the sum over $\pert{\edit} \notin \powerset{\pert{\edit}^\star}$ and upper-bound the sum over 
$\edit \notin \powerset{\edit^\star}$ to obtain a lower bound that is independent of $h$. 

\begin{theorem} \label{thm:smooth-prob-lb}
  For any pair of inputs $\pert{\vec{x}}, \vec{x} \in \Xspace$ we have
  \begin{equation}
      \rho(\pert{\vec{x}}, \vec{x}, \mu_y) \geq \tilde{\rho}(\pert{\vec{x}}, \vec{x}, \mu_y)
      = p_\del^{\abs{\pert{\vec{x}}} - \abs{\vec{x}}} \left( 
          \mu_y - 1 + p_\del^{\frac{1}{2}(\dist_{\LCS}(\pert{\vec{x}}, \vec{x}) + \abs{\vec{x}} - \abs{\pert{\vec{x}}})} 
      \right). \label{eqn:smooth-prob-lb}
  \end{equation}
  where $\dist_{\mathrm{LCS}}(\pert{\vec{x}}, \vec{x})$ is the longest common subsequence (LCS) distance\footnote{%
  The LCS distance is equivalent to the generalized edit distance with $O = \{\del, \ins\}$.} between $\pert{\vec{x}}$ and $\vec{x}$.
\end{theorem}

\subsection{Minimizing over the edit distance neighborhood} \label{sec:regional}
In this section, we complete the derivation of our robustness certificate by minimizing the lower bound 
in \eqref{eqn:smooth-prob-lb} over the edit distance neighborhood:
\begin{equation}
  \tilde{\rho}(\vec{x}; \mu_y) = \min_{\pert{\vec{x}} \in \nbhd_\radius(\vec{x})} \tilde{\rho}(\pert{\vec{x}}, \vec{x}, \mu_y).
  \label{eqn:lb-cert}
\end{equation}
We are interested in general edit distance neighborhoods, where the edit ops $O$ used to define the edit distance 
may be constrained. 
For example, if the attacker is capable of performing elementary substitutions and insertions, but not deletions, then 
$O = \{\sub, \ins\}$. 
As a step towards solving \eqref{eqn:lb-cert}, it is therefore useful to express 
$\tilde{\rho}(\pert{\vec{x}}, \vec{x}, \mu_y)$ in terms of edit op counts, as shown below.

\begin{corollary} \label{cor:pointwise-deletion}
    Suppose there exists a sequence of edits from $\pert{\vec{x}}$ to $\vec{x}$ that consists of $n_\sub$ substitutions, 
    $n_\ins$ insertions and $n_\del$ deletions s.t.\  
    $n_\sub + n_\ins + n_\del = \dist_O(\pert{\vec{x}}, \vec{x})$ and $n_\sub, n_\ins, n_\del \geq 0$. 
    Then 
    \begin{equation*}
        \tilde{\rho}(\pert{\vec{x}}, \vec{x}, \mu_y) 
            = p_\del^{n_\del - n_\ins} \left(\mu_y - 1 + p_\del^{n_\sub + n_\ins}\right).
    \end{equation*}
\end{corollary}

This parameterization of the lower bound enables us to re-express \eqref{eqn:lb-cert} as an optimization problem over 
edit ops counts:
\begin{equation}
    \tilde{\rho}(\vec{x}; \mu_y) = \min_{n_\sub, n_\ins, n_\del \in \mathcal{C}_r} p_\del^{n_\del - n_\ins} \left(\mu_y - 1 + p_\del^{n_\sub + n_\ins}\right),
    \label{eqn:lb-cert-edit-counts}
\end{equation}
where $\mathcal{C}_r$ encodes constraints on the set of counts. 
If any number of insertions, deletions or substitutions are allowed, then the edit distance is known as the 
\emph{Levenshtein distance} and $\mathcal{C}_r$ consists of sets of counts that sum to $\radius$. 
We solve the minimization problem for this case below.

\begin{theorem}[Levenshtein distance certificate] \label{thm:lev-cert} 
    A lower bound on the classifier's confidence within the Levenshtein distance neighborhood 
    $\nbhd_\radius(\vec{x})$ (with $O = \{\del, \ins, \sub\})$ is 
    $\tilde{\rho}(\vec{x}; \mu_y) = \mu_y - 1 + p_\del^\radius$. 
    It follows that the classifier is certifiably robust for any Levenshtein distance ball with radius $\radius$ less 
    than or equal to the \emph{certified radius} 
    $\radius^\star = \floor*{ \nicefrac{\log \left(1 + \nu_y(\vec{\eta}) - \mu_y\right)}{\log p_\del} }$.
\end{theorem}

It is straightforward to adapt this result to account for constraints on the edit ops $O$. 
Results for all combinations of edit ops are provided in Table~\ref{tbl:certificate-O}. 

So far in this section we have obtained results that depend on the classifier's confidence $\mu_y$, assuming it 
can be evaluated exactly. 
However since exact evaluation of $\mu_y$ is not feasible in general (see Section~\ref{sec:practicalities}), we extend 
our results to the probabilistic setting, assuming a $1 - \alpha$ lower confidence bound on $\mu_y$ is available. 
This covers the probabilistic certification procedure described in Figure~\ref{alg:estimation-algorithm}.

\begin{corollary} \label{cor:prob-cert}
  Suppose the procedure in Figure~\ref{alg:estimation-algorithm} returns predicted class $\hat{y}$ with certified radius $\radius^\star$. 
  Then an edit distance robustness certificate of radius $\radius \leq \radius^\star$ holds at $\vec{x}$ with 
  probability $1 - \alpha$.
\end{corollary}

\section{Case study: robust malware detection}
\label{sec:evaluation}
We now present a case study on the application of \rsdel\ to malware detection. 
We report on certified accuracy for Levenshtein and Hamming distance threat models. We show that by tuning decision thresholds we can increase certified radii for the malicious class while maintaining accuracy. We also evaluate \rsdel on a range of published malware classifier attacks. 
Due to space constraints, we present the complete study in Appendices~\ref{app-sec:eval}--\ref{app-sec:eval-attack}, 
where we report on training curves, the computational cost of training and certification, certified radii 
normalized by file size, and results where byte edits are chunked by instructions.

\paragraph{Background}
Malware (malicious software) detection is a long standing problem in security where machine learning is playing 
an increasing role~\cite{microsoft2019new,liu2020review,kaspersky2021machine,blackberry2022cylanceai}. 
Inspired by the success of neural networks in other domains, recent work has sought to design neural network models for 
static malware detection which operate on raw binary executables, represented as byte sequences~\cite{raff2018malware,
krcal2018deep,raff2021classifying}. 
While these models have achieved competitive performance, they are vulnerable to adversarial perturbations 
that allow malware to evade detection%
~\cite{kolosnjaji2018adversarial,park2019generation,kreuk2019deceiving,demetrio2019explaining,
suciu2019exploring,demetrio2021functionality,demetrio2021adversarial,lucas2021malware,song2022mab}. 
Our edit distance threat model reflects these attacks in the malware domain where, even though a variety of perturbations with different semantic effects are possible, 
any perturbation can be represented in terms of elementary byte deletions, insertions and substitutions. 
For perspective on the threat model, consider YARA~\cite{yara2022}, a rule-based tool that is widely 
used for static malware analysis in industry. 
Running Nextron System's YARA rule set\footnote{\url{https://valhalla.nextron-systems.com/}} on a sample of 
binaries from the \vtfeed\ dataset (introduced below), we find 83\% of rule matches are triggered by fewer than 
128~bytes. 
This implies most rules can be evaded by editing fewer than 128~bytes---a regime that is covered by our 
certificates in some instances (see Table~\ref{tbl:cr-statistics-small}).
Further background and motivation for the threat model is provided in Appendix~\ref{app-sec:eval}, along with a 
reduction of static malware detection to sequence classification.

\paragraph{Experimental setup} 
We use two Windows malware datasets: \sleipnir\ which is compiled from public sources following 
\citeteam{aldujaili2018adversarial} and \vtfeed\ which is collected from VirusTotal~\cite{lucas2021malware}.
We consider three malware detection models: a model smoothed with our randomized deletion mechanism (\rsdel), a model 
smoothed with the randomized ablation mechanism proposed by \citeteam{levine2020robustness} (\rsabn), and a 
non-smoothed baseline (\ns). 
All of the models are based on a popular CNN architecture called 
MalConv~\cite{raff2018malware}, and are %
evaluated on a held-out test set. 
We emphasize that \rsdel\ is the only model that provides edit distance certificates (general $O$), while 
\rsabn\ provides a Hamming distance certificate ($O = \{\sub\}$). 
We review \rsabn\ in Appendix~\ref{app-sec:rs-abn}, where we describe modifications required for discrete sequence 
classification, and provide an analysis comparing the Hamming distance certificates of \rsabn\ and \rsdel.
Details about the datasets, models, training procedure, calibration, parameter settings and hardware are 
provided in Appendix~\ref{app-sec:eval-setup}.

\paragraph{Accuracy\slash robustness tradeoffs}

Our first set of experiments investigate tradeoffs between malware detection accuracy and robustness guarantees as 
parameters associated with the smoothed models are varied. 
Table~\ref{tbl:cr-statistics-small} reports clean accuracy, median certified radius (CR) and median certified radius 
normalized by file size (NCR) on the test set for \rsdel\ as a function of $p_\del$. 
A reasonable tradeoff is observed at $p_\del = 99.5\%$ for \sleipnir, where a median certified radius of 
137~bytes is attained with only a 2-3\% drop in clean accuracy from the \ns\ baseline. 
The corresponding median NCR is 0.06\% and varies in the range 0--9\% across the test set.
We also vary the decision threshold $\vec{\eta}$ at $p_\del = 99.5\%$ for \sleipnir, and obtain asymmetrical 
robustness guarantees with a median certified radius up to 582~bytes for the malicious class, while maintaining the 
same accuracies (see Table~\ref{tbl:cr-statistics-smoothprob} in Appendix~\ref{app-sec:eval-cert-lev-dist}). 

Since there are no baseline methods that support edit distance  certificates, we compare with \rsabn, which 
produces a limited Hamming distance certificate. 
Figure~\ref{fig:ca-hamming-sleipnir-cmp} plots the certified accuracy curves for \rsdel\ and \rsabn\ for different 
values of the associated smoothing parameters $p_\del$ and $p_\ab$.
The certified accuracy is the fraction of instances in the test set for which the model's prediction is correct 
\emph{and} certifiably robust at radius $\radius$, and is therefore sensitive to both robustness and accuracy. 
We find that \rsdel\ outperforms \rsabn\ in terms of certified accuracy at all radii $\radius$ when $p_\del = p_\ab$, 
while covering a larger set of inputs (since \rsdel's edit distance ball includes \rsabn's Hamming ball).
Further results and interpretation for these experiments are provided in Appendix~\ref{app-sec:eval-certificate}.

\begin{figure}
  \begin{minipage}[b]{.49\textwidth}
    \centering
    \begin{center}
      \small
      \begin{NiceTabular}{
          l@{}r
          r
          r
          r
        }[colortbl-like]
        \toprule
                                &                  & Clean            & Median        & Median           \\
        Model                   & $p_\del$         & Accuracy         & CR            & NCR \%           \\
        \midrule
        \multicolumn{5}{c}{\rowcolor{gray!10} \sleipnir\ dataset}                                        \\
        \midrule
        \ns                     &  ---             & 98.9\%           & ---           & ---              \\
        \midrule
        \multirow{6}{*}{\rsdel} & 90\%             & 97.1\%           & 6             & 0.0023           \\
                                & 95\%             & 97.8\%           & 13            & 0.0052           \\
                                & 97\%             & 97.4\%           & 22            & 0.0093           \\
                                & 99\%             & 98.1\%           & 68            & 0.0262           \\
                                & \boldcell 99.5\% & \boldcell 96.5\% & \boldcell 137 & \boldcell 0.0555 \\
                                & 99.9\%           & 83.7\%           & 688           & 0.2269           \\
        \midrule
        \multicolumn{5}{c}{\rowcolor{gray!10} \vtfeed\ dataset}                                          \\
        \midrule
        \ns                     & ---              & 98.9\%           & ---           & ---              \\
        \midrule
        \multirow{2}{*}{\rsdel} & 97\%             & 92.1\%           & 22            & 0.0045           \\
                                & 99\%             & 86.9\%           & 68            & 0.0122           \\
        \bottomrule
      \end{NiceTabular}
    \end{center}
    \captionof{table}{Clean accuracy and robustness metrics for \rsdel\ as a function of dataset and deletion 
      probability $p_\del$.
      All metrics are computed on the test set.
      ``Median CR'' is the median certified Levenshtein distance radius in bytes and ``median NCR \%'' is the median 
      certified Levenshtein distance radius normalized as a percentage of the file size.
      A good tradeoff is achieved when $p_\del = 99.5\%$ (in bold).
    }
    \label{tbl:cr-statistics-small}
  \end{minipage}
  \hfill
  \begin{minipage}[b]{0.49\textwidth}
    \centering
    \includegraphics[width=\linewidth]{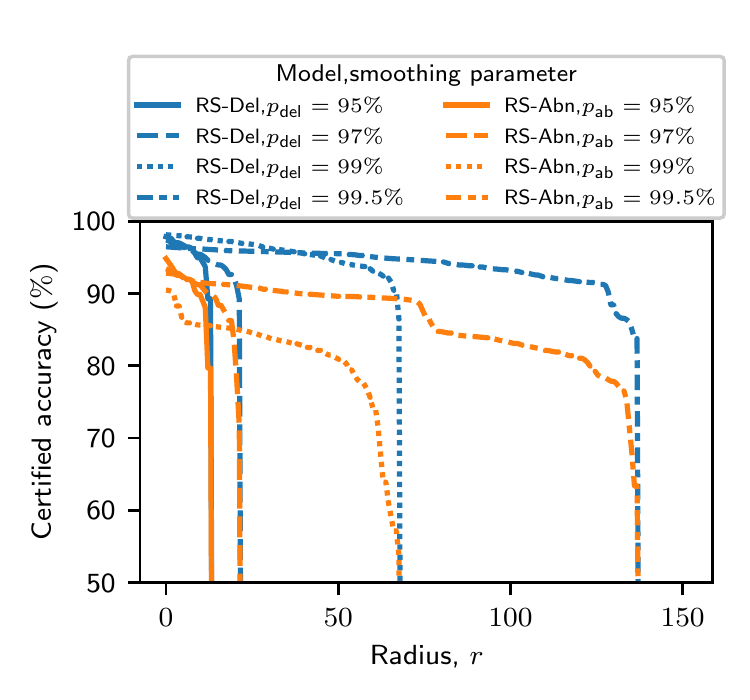}
    \captionof{figure}{
      Certified accuracy for \rsdel\ and \rsabn~\cite{levine2020robustness} as a function of certificate radius 
      (horizontal axis) and strength of the smoothing parameter (line style).
      Results are plotted for the \sleipnir\ test set.
      While \rsdel\ provides a Levenshtein distance certificate
      (with $O = \{\del, \ins, \sub\}$), \rsabn\ provides a more limited Hamming distance certificate ($O = \{\sub\}$).
      The non-smoothed, non-certified model (\ns) achieves a clean accuracy of 98\% in this setting.
    }
    \label{fig:ca-hamming-sleipnir-cmp}
  \end{minipage}
\end{figure}

\paragraph{Empirical robustness to attacks}
Our edit distance certificates are conservative and may underestimate robustness to adversaries with additional 
constraints (e.g., maintaining executability, preserving a malicious payload, etc.). 
To provide a more complete picture of robustness, we subject \rsdel\ and \ns\ to six recently published attacks 
\cite{kreuk2019deceiving,demetrio2019explaining,lucas2021malware,nisi2021lost,demetrio2021functionality} covering 
white-box and black-box settings.
{We adapt gradient-based white-box attacks~\cite{kreuk2019deceiving,lucas2021malware} for randomized smoothing in Appendix~\ref{app-sec:adapt-wb}.} 
We do not constrain the number of edits each attack can make, which yields adversarial examples \emph{well outside} 
the edit distance we can certify for four of six attacks (see Table~\ref{tbl:adv-attacks} of 
Appendix~\ref{app-sec:eval-attack}). 
We measure robustness in terms of the \emph{attack success rate}, defined as the fraction of instances in the 
evaluation set for which the attack flips the prediction from malicious to benign on at least one repeat (we repeat 
each attack five times). 
For both \sleipnir\ and \vtfeed, we observe that \rsdel\ achieves the lowest attack success rate (best robustness) 
for four of six attacks. 
In particular, we observe a 20~percentage point decrease in the success rate of \disp---an attack with no known 
defense~\cite{lucas2021malware}. 
We refer the reader to Appendix~\ref{app-sec:eval-attack} for details of this experiment, including  
setup, results and discussion.

\section{Related work}
\label{sec:related}

There is a rich body of research 
on certifications under $\ell_p$-norm-bounded threat models~\cite{wong2018provable,dvijotham2018training,raghunathan2018certified,weng2018towards,
mirman2018differentiable,tsuzuku2018lipschitz,gowal2019effectiveness,lecuyer2019certified,cohen2019certified,zhang2020towards,leino2021globally,jia2022almost,hammoudeh2023feature}.
While a useful abstraction for computer vision, such certifications are inadequate for many problems including perturbations to executable files considered in this work.
Even in computer vision, $\ell_p$-norm bounded defenses can be circumvented by image translation,
rotation, blur, and other human-imperceptible transformations that induce extremely large $\ell_p$ distances.
One solution is to re-parametrize the norm-bounded distance
in terms of image transformation parameters \cite{fischer2020certified,li2021tss,hao2022gssmooth}.
NLP faces a different issue:
while the $\ell_0$ threat model covers adversarial word substitution~\cite{ren2019generating,zeng2023certified}, it is too broad and covers many natural (non-adversarial) examples as well.
For example, ``He loves cat'' and ``He hates cat'' are 
 1~word in $\ell_0$ distance from ``He likes cat'', but are semantically different.
A radius 1 certificate will force a wrong prediction for at least one neighbor.
To address this, \citeteam{jia2019certified} and \citeteam{ye2020safer} constrain the threat model
to synonyms only.

In this paper we go beyond the $\ell_0$ word substitution threat model of previous work~\cite{lee2019tight,
levine2020robustness,jia2022almost,hammoudeh2023feature}, as
consideration of insertions and deletions is necessary in domains such as malware analysis. Such edits are not captured
by the $\ell_0$ threat model: there is no fixed input size, and even when edits are size-preserving, a few edits may
lead to large $\ell_0$ distances.
Arguably, our edit distance threat model for sequences and \rsdel\ mechanism are of independent interest to natural 
language also. 

Certification has been studied for variations of edit distance defined for sets and graphs. 
\citeteam{liu2021pointguard} apply randomized smoothing using a subsampling mechanism to certify point cloud 
classifiers against edits that include insertion, deletion and substitution of points. 
Since a point cloud is an \emph{unordered} set, the edit distance takes a simpler form than for sequences---it can be 
expressed in terms of set cardinalities rather than as a cost minimization problem over edit paths.
This simplifies the analysis, allowing \citeauthor{liu2021pointguard} to obtain a tight certificate via the 
Neyman-Pearson lemma, which is not feasible for sequences. 
In parallel work, \citeteam{schuchardt2023adversarial} consider edit distance certification for graph classification, 
as an application of a broader certification framework for group equivariant tasks. 
They apply sparsity-aware smoothing~\cite{bojchevski2020efficient} to an isomorphism equivariant base classifier, to 
yield a smoothed classifier that is certifiably robust under insertions\slash deletions of edges and node attributes.

Numerous empirical defense methods have been proposed to improve robustness of machine learning 
classifiers in security systems~\cite{demontis2019yes,quiring2020against,saha2023adversarial,chen2021learning,chen2020training}. 
\citeteam{romeo2018adversarially} and \citeteam{chen2021learning}
develop classifiers that are verifiably robust if their manually crafted features conform to particular
properties (e.g., monotonicity, stability).
These approaches permit a combination of (potentially vulnerable) learned behavior with domain knowledge,
and thereby aim to mitigate adversarial examples. 
\citeteam{chen2020training} seek guarantees against subtree insertion and deletion for PDF malware classification. Using convex over-approximation~\cite{wong2018provable,wong2018scaling} previously applied to computer vision, they certify fixed-input dimension classifiers popular in PDF malware analysis. 
Concurrent to our work, \citeteam{saha2023adversarial} propose to certify classifiers against patch-based
attacks by aggregating predictions of fixed-sized chunks of input binaries. 
The patch attack threat model, however, is not widely assumed in the evasion literature for malware detectors and
can be readily broken by many published attacks \cite{lucas2021malware,kreuk2019deceiving}.
Moreover, their de-randomized smoothing design assumes a fixed-width input (via padding and/or trimming) and reduces patch-based attacks to gradient-based $\ell_p$ attacks. 
While tight analysis exists for arbitrary randomized smoothing mechanisms \cite{lee2019tight}, they are computationally infeasible with the edit distance threat model.
Overall, we are the first to explore certified adversarial defenses that apply to sequence classifiers under the edit distance
threat model.

\section{Conclusion}
\label{sec:conclusion}

In this paper, we study certified robustness of discrete sequence classifiers.
We identify critical limitations of the $\ell_p$-norm bounded threat model in sequence classification
and propose edit distance-based robustness, covering substitution, deletion and insertion perturbations.
We then propose a novel deletion smoothing mechanism called \rsdel\ that is equipped with certified guarantees
under several constraints on the edit operations. 
We present a case study of \rsdel\ and its certifications applied to malware analysis. We consider two malware datasets using a recent static deep malware detector, MalConv~\cite{raff2018malware}.
We find that \rsdel\ can certify radii as large as $128$ bytes (in Levenshtein distance) without significant loss in detection accuracy. 
A certificate of this size covers in excess of $10^{606}$ files in the proximity of a 10KB input file (see Appendix~\ref{app-sec:exact}).
Results also demonstrate \rsdel\ improving robustness against published attacks well beyond the certified radius.

\paragraph{Broader impact and limitations} Robustness certification seeks to quantify the risk of adversarial examples while randomized smoothing both enables certification and acts
to mitigate the impact of attacks. Randomized smoothing can degrade (benign) accuracy of undefended models as demonstrated in our results at higher
smoothing levels. 
While we have strived to select high quality datasets for our case study, we note that accuracy-robustness tradeoffs may vary for different datasets and\slash or model architectures.
Our approach is scalable relative to alternative certification strategies, however it does incur computational overheads.
Finally, it is known that randomized smoothing can have disparate impacts on class-wise accuracy~\cite{mohapatra21hidden}.

\begin{acks}
This work was supported by the Department of Industry, Science, and 
Resources, Australia under AUSMURI CATCH, and the U.S.\ Army Research Office 
under MURI Grant W911NF2110317.
\end{acks}

\bibliographystyle{unsrtnat}
\bibliography{malware}

\newpage
\appendix

\section{Brute-force edit distance certification}
\label{app-sec:exact}

In this appendix, we show that an edit distance certification mechanism based on brute-force search is 
computationally infeasible. 
Suppose we are interested in issuing an edit distance certificate at radius $\radius$ for a sequence classifier $f$ 
at input $\vec{x}$. 
Recall from \eqref{eqn:robustness-cert} that in order to issue a certificate, we must show there 
exists no input $\pert{\vec{x}}$ within the edit distance neighborhood $\nbhd_\radius(\vec{x})$ that would change 
$f$'s prediction. 
This problem can theoretically be tackled in a brute-force manner, by querying $f$ for all inputs in 
$\nbhd_\radius(\vec{x})$. 
In the best case, this would take time linear in $|\nbhd_\radius(\vec{x})|$, assuming $f$ responds to 
queries in constant time. 
However the following lower bound \cite{charalampopoulos2020unary}, shows that the size of the edit distance 
neighborhood is too large even in the best case:
\begin{equation*}
  |\nbhd_\radius(\vec{x})| \geq \sum_{i = 0}^{\radius} 255^i \sum_{j = i - \radius}^{\radius} \binom{\abs{\vec{x}} + j}{i} \geq 255^r.
\end{equation*}
For example, applying the loosest bound that is independent of $\vec{x}$, we see that brute-force certification at 
radius $\radius = 128$ would require in excess of $255^\radius \approx 10^{308}$ queries to $f$. 
In contrast, our probabilistic certification mechanism (Figure~\ref{alg:estimation-algorithm}) makes 
$n_\mathrm{pred} + n_\mathrm{bnd}$ queries to $f$, and we can provide high probability guarantees when 
the number of queries is of order $10^3$ or $10^4$.

\section{Proofs for Section~\ref{sec:certificate}} \label{app-sec:proofs}

In this appendix, we provide proofs of the theoretical results stated in Section~\ref{sec:certificate}. 

\subsection{Proof of Proposition~\ref{prop:cert-suff}}
    A sufficient condition for \eqref{eqn:rsdel-cert} is 
    \begin{equation}
      \min_{\pert{\vec{x}} \in \nbhd_\radius(\vec{x})} \min_{h \in \Fspace(\vec{x})} p_y(\pert{\vec{x}}; h) 
        \geq \max_{\pert{\vec{x}} \in \nbhd_\radius(\vec{x})} \max_{h \in \Fspace(\vec{x})} 
          \left(\eta_y + \max_{y' \neq y} p_{y'}(\pert{\vec{x}}; h) -  \min_{y' \neq y} \eta_{y'} \right).
      \label{eqn:rsdel-cert-inequal-suff}
    \end{equation}
  
    We first consider the multi-class case where $\abs{\Yspace} > 2$. 
    If $\eta_y \geq \min_{y' \neq y} \eta_{y'}$, then  
    $p_y(\pert{\vec{x}}; h) \geq \max_{y' \neq y} p_{y'}(\pert{\vec{x}}; h)$ by \eqref{eqn:rsdel-cert-inequal-suff} 
    and we can upper-bound $\max_{y' \neq y} p_{y'}(\pert{\vec{x}}; h)$ by $\frac{1}{2}$. 
    On the other hand, if $\eta_y \geq \min_{y' \neq y} \eta_{y'}$, we can only upper-bound 
    $\max_{y' \neq y} p_{y'}(\pert{\vec{x}}; h)$ by $1$. 
    Thus when $\abs{\Yspace} > 2$ \eqref{eqn:rsdel-cert-inequal-suff} implies
    \begin{equation*}
      \min_{\pert{\vec{x}} \in \nbhd_\radius(\vec{x})} \min_{h \in \Fspace(\vec{x})} p_y(\pert{\vec{x}}; h) 
        \geq \begin{cases}
        \frac{1}{2} + \eta_y - \min_{y' \neq y} \eta_{y'}, 
          & \eta_y \geq \min_{y' \neq y} \eta_{y'}, \\
        1 + \eta_y - \min_{y' \neq y} \eta_{y'}, 
          & \eta_y < \min_{y' \neq y} \eta_{y'}.
      \end{cases}
    \end{equation*}
  
    Next, we consider the binary case where $\abs{\Yspace} = 2$. 
    Since the confidences sum to 1, we have $\max_{y' \neq y} p_{y'}(\pert{\vec{x}}; h) = 1 - p_y(\pert{\vec{x}}; h)$. 
    Putting this in \eqref{eqn:rsdel-cert-inequal-suff} implies  
    \begin{equation*}
      \min_{\pert{\vec{x}} \in \nbhd_\radius(\vec{x})} \min_{h \in \Fspace(\vec{x})} p_y(\pert{\vec{x}}; h) 
        \geq \frac{1 + \eta_y - \min_{y' \neq y} \eta_{y'}}{2}.
    \end{equation*}  

\subsection{Proof of Lemma~\ref{lem:equiv-edits}}

    Let $r_S \colon S \to \{1, \ldots, \abs{S}\}$ be a bijection that returns the \emph{rank} of an element in an 
    ordered set $S$. 
    Let $\dot{r}_S \colon \powerset{S} \to \powerset{\{1, \ldots, \abs{S}\}}$ be an elementwise extension of $r_S$ 
    that returns a \emph{set of ranks} for an ordered set of elements---i.e., $\dot{r}_S(U) = \{\, r_S(i) : i \in U \,\}$ 
    for $U \subseteq S$.
    We claim $m(\pert{\edit}) = \dot{r}_{\edit^\star}^{-1}(\dot{r}_{\pert{\edit}^\star}(\pert{\edit}))$ is a bijection that 
    satisfies the required property.
    
    To prove the claim, we note that $m$ is a bijection from $\powerset{\pert{\edit}^\star}$ to 
    $\powerset{\edit^\star}$ since it is a composition of bijections 
    $\dot{r}_{\pert{\edit}^\star}: \powerset{\pert{\edit}^\star} \to \powerset{\{1, \ldots, l\}}$ and 
    $\dot{r}_{\edit^\star}^{-1}: \powerset{\{1, \ldots, l\}} \to \powerset{\edit^\star}$ where 
    $l = \abs{\pert{\edit}^\star} = \abs{\edit^\star}$. 
    Next, we observe that $\dot{r}_{\pert{\edit}^\star}(\pert{\edit})$ relabels indices in $\pert{\edit}$ so they have the same effect 
    when applied to $\vec{z}^\star$ as $\pert{\edit}$ on $\pert{\vec{x}}$ (this also holds for $\dot{r}_{\edit^\star}$ and 
    $\edit$).
    Thus 
    \begin{align*}
        \apply(\pert{\vec{x}}, \pert{\edit}) 
        &= \apply(\vec{z}^\star, \dot{r}_{\pert{\edit}^\star}(\pert{\edit})) \\
        &= \apply(\vec{z}^\star, \dot{r}_{\edit^\star}(\dot{r}_{\edit}^{-1} (\dot{r}_{\pert{\edit}^\star}(\pert{\edit})))) \\
        &= \apply(\vec{x}, m(\pert{\edit}))
    \end{align*}
    as required.
    To prove the final statement, we use \eqref{eqn:edit-distribution}, \eqref{eqn:apply-edit} and 
    \eqref{eqn:smoothed-prob-summand} to write
    \begin{align*}
        \frac{s(\pert{\edit}, \pert{\vec{x}}; h)}{s(\edit, \vec{x}; h)} 
        &= \frac{\ind{h(\apply(\pert{\vec{x}}, \pert{\edit})) = y} 
                p_\del^{\abs{\pert{\vec{x}}} - \abs{\pert{\edit}}} (1 - p_\del)^{\abs{\pert{\edit}}}}
            {\ind{h(\apply(\vec{x}, \edit)) = y} 
                p_\del^{\abs{\vec{x}} - \abs{\edit}} (1 - p_\del)^{\abs{\edit}}} \\
        &= \frac{p_\del^{\abs{\pert{\vec{x}}} - \abs{\vec{z}}} (1 - p_\del)^{\abs{\vec{z}}} \ind{h(\vec{z}) = y}}
            {p_\del^{\abs{\vec{x}} - \abs{\vec{z}}} (1 - p_\del)^{\abs{\vec{z}}} \ind{h(\vec{z}) = y}} \\ 
        &= p_\del^{\abs{\pert{\vec{x}}} - \abs{\vec{x}}},
    \end{align*}
    where the second last line follows from the fact that 
    $\apply(\pert{\vec{x}}, \pert{\edit}) = \apply(\vec{x}, \edit) = \vec{z}$.

\subsection{Proof of Theorem~\ref{thm:smooth-prob-lb}}

    Let $\pert{\edit}^\star$ and $\edit^\star$ be defined as in Lemma~\ref{lem:equiv-edits}.
    We derive an upper bound on the sum over $\edit \in \powerset{\edit^\star}$ that appears 
    in~\eqref{eqn:smooth-prob-mu}.
    Observe that 
    \begin{align}
        \sum_{\edit \notin \powerset{\edit^\star}} s(\edit, \vec{x}; h) 
            &\leq \sum_{\edit \notin \powerset{\edit^\star}} 
                \Pr\left[G(\vec{x}) = \edit\right] \nonumber \\
            &= 1 - \sum_{\edit \in \powerset{\edit^\star}} 
                \Pr\left[G(\vec{x}) = \edit\right] \nonumber \\
            &= 1 - p_\del^{\abs{\vec{x}} - \abs{\edit^\star}} 
                \sum_{\abs{\edit} = 0}^{\abs{\edit^\star}} \binom{\abs{\edit^\star}}{\abs{\edit}}
                    p_\del^{\abs{\edit^\star} - \abs{\edit}} (1 - p_\del)^{\abs{\edit}} \nonumber \\
            &= 1 - p_\del^{\abs{\vec{x}} - \abs{\edit^\star}}, \label{eqn:overlap}
    \end{align}
    where the first line follows from the inequality $\ind{h(\apply(\vec{x}, \edit) = y)} \leq 1$; 
    the second line follows from the law of total probability; the third line follows by constraining the indices 
    $\{1, \ldots, \abs{\vec{x}} \} \setminus \pert{\edit}^\star$ to be deleted; and the last line follows from the 
    normalization of the binomial distribution.
    Putting \eqref{eqn:overlap} and $\sum_{\pert{\edit} \in 2^{\pert{\edit}^\star}} s(\pert{\edit}, \pert{\vec{x}}; h) \geq 0$ in 
    \eqref{eqn:smooth-prob-mu} gives
    \begin{align}
        p_y(\pert{\vec{x}}; h) &\geq p_\del^{\abs{\pert{\vec{x}}} - \abs{\vec{x}}} \left(
            \mu_y - 1 - p_\del^{\abs{\vec{x}} - \abs{\edit^\star}}
        \right) \nonumber \\
        &= p_\del^{\abs{\pert{\vec{x}}} - \abs{\vec{x}}} \left(
            \mu_y - 1 - p_\del^{\frac{1}{2} (\dist_{\LCS}(\pert{\vec{x}}, \vec{x}) + \abs{\vec{x}} - \abs{\pert{\vec{x}}})}
        \right). \label{eqn:bound-lcs}
    \end{align}
    In the second line above we use the following relationship between the LCS distance and the length of the 
    LCS $\abs{\vec{z}^\star} = \abs{\edit^\star}$:
    \begin{equation*}
        \dist_{\LCS}(\pert{\vec{x}}, \vec{x}) = \abs{\pert{\vec{x}}} + \abs{\vec{x}} - 2 \abs{\vec{z}^\star}.
    \end{equation*}
    Since \eqref{eqn:bound-lcs} is independent of the base classifier $h$, the lower bound on 
    $\rho(\pert{\vec{x}}, \vec{x}, \mu_y)$ follows immediately.

\subsection{Proof of Corollary~\ref{cor:pointwise-deletion}}

    Since the length of $\vec{x}$ can only be changed by inserting or deleting elements in $\pert{\vec{x}}$, we have 
    \begin{equation}
        \abs{\vec{x}} - \abs{\pert{\vec{x}}} = n_\ins - n_\del. \label{eqn:length-diff-counts}
    \end{equation}
    We also observe that the LCS distance can be uniquely decomposed in terms of the counts of insertion ops 
    $m_\ins$ and deletion ops $m_\del$: $\dist_{\LCS}(\pert{\vec{x}}, \vec{x}) = m_\del + m_\ins$. 
    These counts can in turn be related to the given decomposition of edit ops counts for generalized edit distance. 
    In particular, any substitution must be expressed as an insertion and deletion under LCS distance, which 
    implies $m_\ins = n_\ins + n_\sub$ and $m_\del = n_\del + n_\sub$.
    Thus we have
    \begin{equation}
        \dist_{\LCS}(\pert{\vec{x}}, \vec{x}) = n_\del + n_\ins + 2 n_\sub. \label{eqn:lcs-dist-counts} 
    \end{equation}
    Substituting \eqref{eqn:length-diff-counts} and \eqref{eqn:lcs-dist-counts} in \eqref{eqn:smooth-prob-lb} gives 
    the required result.

\subsection{Proof of Theorem~\ref{thm:lev-cert}}

    Eliminating $n_\sub$ from \eqref{eqn:lb-cert-edit-counts} using the constraint
    $n_\sub = \radius - n_\del - n_\ins$, we obtain a minimization problem in two variables:
    \begin{equation*}
        \begin{aligned}
            &\!\min_{n_\ins, n_\del \in \mathbb{N}_0}  && \psi(n_\ins, n_\del) \\
            &\text{s.t.}                               && 0 \leq n_\ins + n_\del \leq \radius  
        \end{aligned}
    \end{equation*}
    where $\psi(n_\ins, n_\del) = p_\del^{n_\del - n_\ins} \left(\mu_y - 1 + p_\del^{r - n_\del}\right)$.
    Observe that $\psi$ is monotonically increasing in $n_\ins$ and $n_\del$:
    \begin{align*}
        \frac{\psi(n_\ins + 1, n_\del)}{\psi(n_\ins, n_\del)} 
        &= \frac{1}{p_\del} 
        \geq 1 \\
        \frac{\psi(n_\ins, n_\del + 1)}{\psi(n_\ins, n_\del)} 
        &= \frac{(\mu_y - 1) p_\del^{n_\del + 1} + p_\del^{\radius}}{(\mu_y - 1) p_\del^{n_\del} + p_\del^{\radius}}\geq 1,
    \end{align*}
    where the second inequality follows since we only consider $\radius$ and $\mu_y$ such that the numerator and 
    denominator are positive.
    Thus the minimizer is $(n_\ins^\star, n_\del^\star, n_\sub^\star) = (0, 0, \radius)$ and we find 
    $\rho(\vec{x}; \mu_y) = \mu_y - 1 + p_\del^\radius$. 
    The expression for the certified radius follows by solving 
    $\rho(\vec{x}; \mu_y) \geq \nu_y(\vec{\eta})$ for non-negative integer $r$.

\subsection{Proof of Corollary~\ref{cor:prob-cert}}

    Recall that Corollary~\ref{cor:pointwise-deletion} gives the following lower bound on the classifier's confidence 
    at $\vec{x}$:
    \begin{equation*}
        \tilde{\rho}(\pert{\vec{x}}, \vec{x}, \mu_y) = p_\del^{n_\del - n_\ins} \left(\mu_y - 1 + p_\del^{n_\sub + n_\ins} \right).
    \end{equation*}
    Observe that we can replace $\mu_y$ by a lower bound $\umu_y$ that holds with probability $1 - \alpha$ 
    (as is done in lines 4--6 of Figure~\ref{alg:estimation-algorithm}) and obtain a looser lower bound 
    $\tilde{\rho}(\pert{\vec{x}}, \vec{x}, \umu_y) \leq \tilde{\rho}(\pert{\vec{x}}, \vec{x}, \mu_y)$ that holds with probability 
    $1 - \alpha$. 
    Crucially, this looser lower bound has the same functional form, so all results depending on 
    Corollary~\ref{cor:pointwise-deletion}, namely Theorem~\ref{thm:lev-cert} and Table~\ref{tbl:certificate-O}, 
    continue to hold albeit with probability $1 - \alpha$.

\section{Background for malware detection case study} \label{app-sec:eval}
In this appendix, we provide background for our case study on malware detection, including motivation for studying 
certified robustness of malware detectors, a formulation of malware detection as a sequence classification problem, and 
a threat model for adversarial examples.

\subsection{Motivation}
Malware (malicious software) detection is a vital capability for proactively defending against 
cyberattacks. %
Despite decades of progress, building and maintaining effective malware detection systems remains a challenge, 
as malware authors continually evolve their tactics to bypass detection and exploit new vulnerabilities. 
One technology that has lead to advancements in malware detection, is the application of machine learning (ML), 
which is now used in many commercial systems~\cite{tsao2019faster,microsoft2019new,kaspersky2021machine,
blackberry2022cylanceai} and continues to be an area of interest in the malware research community 
\cite{vinayakumar2019robust,aghakhani2020when,liu2020review,raff2021classifying}. 
While traditional detection techniques rely on manually-curated signatures or detection rules, ML allows a 
detection model to be learned from a training corpus, that can potentially generalize to unseen programs.

Although ML has an apparent advantage in detecting previously unseen malware, recent research has shown that ML-based 
static malware detectors can be evaded by applying adversarial perturbations~\cite{kolosnjaji2018adversarial,
park2019generation,kreuk2019deceiving,demetrio2019explaining,suciu2019exploring,demetrio2021functionality,
demetrio2021adversarial,lucas2021malware,song2022mab}. 
A variety of perturbations have been considered with different effects at the semantic level, however all of them
can be modeled as inserting, deleting and\slash or substituting bytes. 
This prompts us to advance certified robustness for sequence classifiers within this general threat
model---where an attacker can perform byte-level edits.

\subsection{Related work}

Several empirical defense methods have been proposed to improve robustness of ML classifiers~\cite{demontis2019yes, quiring2020against}. 
\citeteam{romeo2018adversarially} compose manually crafted Boolean features with a classifier that is constrained to be monotonically increasing with respect to selected inputs. 
This approach permits a combination of (potentially vulnerable) learned behavior with domain knowledge, and thereby aims to mitigate adversarial examples. 
\citeteam{demontis2019yes} show that the sensitivity of linear support vector machines to adversarial perturbations
can be reduced by training with $\ell_\infty$ regularization of weights.
In another work, \citeteam{quiring2020against} take advantage of
heuristic-based semantic gap detectors and an ensemble of feature classifiers to improve empirical robustness. 
Compared to our work on certified adversarial defenses, these approaches do not provide formal guarantees.

\emph{Binary normalization}
\cite{perriot2003defeating,christodorescu2005malware,walenstein2006normalizing,bruschi2007code}
was originally proposed to defend against polymorphic\slash metamorphic malware, and
can also be seen as a mitigation to certain adversarial examples. 
It attempts to sanitize binary obfuscation techniques by mapping malware to a canonical
form before running a detection algorithm.
However, binary normalization cannot fully mitigate attacks like \disp\ (see Table~\ref{tbl:adv-attacks}), as deducing opaque and evasive
predicates are NP-hard problems~\cite{lucas2021malware}.

Dynamic analysis can provide additional insights for malware detection.
In particular, it can record a program's  behavior while executing it in a sandbox (e.g., collecting a call graph or network traffic) 
\cite{rieck2008learning,ye2008sbmds,qiao2013analyzing,jiang2018dlgraph,zhang2020dynamic}. 
Though detectors built on top of dynamic analysis can be more difficult to evade,
as the attacker needs to obfuscate the program's behavior,
they are still susceptible to adversarial perturbations.
For example, an attacker may insert API calls to obfuscate a malware's behavior~\cite{rosenbery2018generic,hu2018black,rosenberg2020query,fadadu2019evading}. 
Applying \rsdel\ to certify detectors that operate on call sequences \cite{zhang2020dynamic} or more general 
 dynamic features would be an interesting future direction.

\subsection{Static ML-based malware detection} \label{sec:malware-detection}
We formulate malware detection as a sequence classification problem, where the objective is to classify a file 
in its raw byte sequence representation as malicious or benign. 
In the notation of Section~\ref{sec:formulation}, we assume the space of input sequences (files) is 
$\Xspace = \Omega^\star$ where $\Omega = \{0, 1, \ldots, 255\}$ denotes the set of bytes, and we assume the set of
classes is $\Yspace = \{0, 1\}$ where 1 denotes the `malicious' class and 0 denotes the `benign' class.
Within this context, a \emph{malware detector} is simply a classifier $f: \Xspace \to \Yspace$. 

\paragraph{Detector assumptions}
Malware detectors are often categorized according to whether they perform static or dynamic analysis. 
Static analysis extracts information without executing code, whereas dynamic analysis extracts information by executing 
code and monitoring its behavior. 
In this work, we focus on machine learning-based static malware detectors, where the ability to extract and synthesize 
information is learned from data. 
Such detectors are suitable as base classifiers for \rsdel, as they can learn to make (weak) predictions for incomplete 
files where chunks of bytes are arbitrarily removed. 
We note that dynamic malware detectors are not compatible with \rsdel, since it is not generally possible to execute 
an incomplete file.

\paragraph{Incorporating semantics} \label{app-sec:exploit-semantics}
In Section~\ref{sec:practicalities}, we noted that our methods are compatible with \emph{sequence chunking} where 
the original input sequence is partitioned into chunks, and reinterpreted as a sequence of chunks rather than a 
sequence of lower-level elements. 
In the context of malware detection, we can partition a byte sequence into semantically meaningful chunks using 
information from a disassembler, such as Ghidra~\cite{ghidra}. 
For example, a disassembler can be applied to a Windows executable to identify chunks of raw bytes that correspond to 
components of the header, machine instructions, raw data, padding etc. 
Applying our deletion smoothing mechanism at the level of semantic chunks, rather than raw bytes, may improve 
robustness as it excludes edits within chunks that may be semantically invalid. 
It also yields a different chunk-level edit distance certificate, that may cover a larger set of adversarial examples 
than a byte-level certificate of the same radius. 
Figure~\ref{fig:insn-del-exp} illustrates the difference between byte-level and chunk-level deletion for a 
Windows executable, where chunks correspond to machine instructions (such as \texttt{push ebp}) or non-instructions 
(\texttt{NI}).

\begin{figure*}
    \centering{\scriptsize
    \hfill
    \begin{tabular}[t]{>{\ttfamily}l >{\ttfamily}l *{2}{>{\ttfamily}l}}
        \\
        \multicolumn{4}{c}{\bf{Original file}}                                                                                                                      \\[0.5ex]
        \multicolumn{1}{l}{File offset} & \multicolumn{1}{l}{Byte}      & \multicolumn{2}{l}{Instruction chunks}                                                    \\\hhline{*{4}{-}}
        00000000                        & 77                            & \multicolumn{2}{>{\ttfamily}l}{\cellcolor{gray!40}NI}                                     \\\hhline{*{4}{-}}
        00000001                        & \cellcolor{gray!20}90         & \multicolumn{2}{>{\ttfamily}l}{NI}                                                        \\\hhline{*{4}{-}}
        00000002                        & 144                           & \multicolumn{2}{>{\ttfamily}l}{\cellcolor{gray!40}NI}                                     \\\hhline{*{4}{-}}
        \myvdots                        & \myvdots                      & \multicolumn{2}{>{\ttfamily}l}{\myvdots}                                                  \\\hhline{*{4}{-}}
        00000400                        & 85                            & \multirow{1}{*}{push}                     & \multirow{1}{*}{ebp}                          \\\hhline{*{4}{-}}
        00000401                        & \cellcolor{gray!20}139        & \multirow{2}{*}{mov}                      & \multirow{2}{*}{ebp, esp}                     \\\hhline{*{2}{-}*{2}{~}}
        00000402                        & \cellcolor{gray!20}236        &                                           &                                               \\\hhline{*{4}{-}}
        \arrayrulecolor{black}
        00000403                        & 131                           & \cellcolor{gray!40}                       & \cellcolor{gray!40}                           \\\hhline{*{2}{-}*{2}{>{\arrayrulecolor{gray!40}}-}}
        \arrayrulecolor{black}
        00000404                        & 236                           & \cellcolor{gray!40}                       & \cellcolor{gray!40}                           \\\hhline{*{2}{-}*{2}{>{\arrayrulecolor{gray!40}}-}}
        \arrayrulecolor{black}
        00000405                        & \cellcolor{gray!20}92         & \multirow{-3}{*}{\cellcolor{gray!40}sub}  & \multirow{-3}{*}{\cellcolor{gray!40}esp, 5Ch} \\\hhline{*{4}{-}}
        \myvdots                        & \myvdots                      & \multicolumn{2}{>{\ttfamily}l}{\myvdots}                                                  \\
    \end{tabular}
    \hfill
    \begin{tabular}[t]{*{2}{>{\ttfamily}l}}
        \multicolumn{2}{c}{\bf{File under}}                             \\
        \multicolumn{2}{c}{\bf{byte-level deletion (\textsc{Byte})}}    \\[0.5ex]
        \multicolumn{1}{l}{File offset} & \multicolumn{1}{l}{Byte}      \\\hline
        00000000                        & 77                            \\\hline
        00000002                        & 144                           \\\hline
        \myvdots                        & \myvdots                      \\\hline
        00000400                        & 85                            \\\hline
        00000403                        & 131                           \\\hline
        00000404                        & 236                           \\\hline
        \myvdots                        & \myvdots                      \\
    \end{tabular}
    \hfill
    \begin{tabular}[t]{*{2}{>{\ttfamily}l}}
        \multicolumn{2}{c}{\bf{File under}}                             \\
        \multicolumn{2}{c}{\bf{chunk-level deletion (\textsc{Insn})}}   \\[0.5ex]
        \multicolumn{1}{l}{File offset} & \multicolumn{1}{l}{Chunk}     \\\hline
        00000001                        & 90                            \\\hline
        \myvdots                        & \myvdots                      \\\hline
        00000400                        & 85                            \\\hline
        00000401                        & 139 236                       \\\hline
        \myvdots                        &                               \\
    \end{tabular}
    \hfill}
    \caption{
        Illustration of the deletion smoothing mechanism applied to an executable file at the byte-level 
        versus chunk-level. 
        \emph{Left:} 
        An executable file where the elementary byte sequence representation is shown in the 2nd column and chunks 
        that correspond to machine instructions are shown in the 3rd column (sourced from the Ghidra~\cite{ghidra} 
        disassembler). 
        Bytes that do not correspond to machine instructions are marked \texttt{NI}. 
        Shading represents bytes (light gray) or instruction chunks (dark gray) that are deleted in the corresponding 
        perturbed file to the right.
        \emph{Middle:}
        A perturbed file produced by the deletion mechanism operating at the byte level (\textsc{Byte}). 
        Notice that individual instructions may be partially deleted.
        \emph{Right:} 
        A perturbed file produced by the deletion mechanism operating at the chunk-level (\textsc{Insn}).}
    \label{fig:insn-del-exp}
\end{figure*}

\subsection{Threat model} \label{sec:threat-model}

We next specify the modeled attacker's goals, capabilities and knowledge for our malware detection case 
study~\cite{barreno2006can}.

\paragraph{Attacker's objective} 
We consider evasion attacks against a malware detector $f: \Xspace \to \Yspace$, where the attacker's objective is 
to transform an executable file $\vec{x}$ so that it is misclassified by $f$. 
To ensure the attacked file $\pert{\vec{x}}$ is useful after evading detection, we require that it is 
\emph{functionally equivalent} to the original file $\vec{x}$. 
We focus on evasion attacks that aim to misclassify a \emph{malicious} file as \emph{benign} in our experiments, 
as these attacks dominate prior work~\cite{demetrio2021adversarial}.  
However, the robustness certificates derived in Section~\ref{sec:certificate} also cover attacks in the opposite 
direction---where a \emph{benign} file is misclassified as \emph{malicious}. 

\paragraph{Attacker's capability}
We measure the attacker's capability in terms of the number of elementary edits they make to the original file
$\vec{x}$. 
If the attacker is capable of making up to $c$ elementary edits, then they can transform $\vec{x}$ into any file in 
the edit distance ball of radius $c$ centred on $\vec{x}$:
\begin{equation*}
  \Aspace_c(\vec{x}) = \{\pert{\vec{x}} \in \Xspace: \dist_{O}(\vec{x}, \pert{\vec{x}}) \leq c\}.
\end{equation*}
Here $\dist_O(\vec{x}, \pert{\vec{x}})$ denotes the edit distance from the original file $\vec{x}$ to the attacked 
file $\pert{\vec{x}}$ under the set of edit operations (ops) $O$. 
We assume $O$ consists of elementary byte-level or chunk-level deletions (\del), insertions (\ins) and substitutions 
(\sub), or a subset of these operations. 

We note that edit distance is a reasonable proxy for the cost of running evasion attacks that iteratively apply 
localized functionality-preserving edits (e.g., \cite{park2019generation,demetrio2019explaining,nisi2021lost,
lucas2021malware,song2022mab}). 
For these attacks, the edit distance scales roughly linearly with the number of attack iterations, and therefore the  
attacker has an incentive to minimize edit distance. 
While attacks do exist that make millions of elementary edits in the malware domain (e.g., \cite{demetrio2021functionality}), 
we believe that an edit distance-constrained threat model is an important step towards realistic threat models 
for certified malware detection. 
(To examine the effect of large edits on robustness we include the \secinj\ 
attack~\cite{demetrio2021functionality} in experiments covered in Appendix~\ref{app-sec:eval-attack}.)

\begin{remark}
  The set $\Aspace_c(\vec{x})$ \emph{overestimates} the capability of an edit distance-constrained attacker, 
  because it may include files that are not functionally equivalent to $\vec{x}$. 
  For example, $\Aspace_c(\vec{x})$ may include files that are not malicious (assuming $\vec{x}$ is malicious) or 
  files that are invalid executables. 
  This poses no problem for certification, since overestimating an attacker's capability merely leads to a stronger 
  certificate than required. 
  Indeed, overestimating the attacker's capability seem necessary, as functionally equivalent files are difficult 
  to specify, let alone analyze.
\end{remark}

\paragraph{Attacker's knowledge} %
In our certification experiments in Appendix~\ref{app-sec:eval-certificate}, we assume the attacker has 
full knowledge of the malware detector and certification scheme. 
When testing published attacks in Appendix~\ref{app-sec:eval-attack}, we consider both white-box and black-box 
access to the malware detector. 
In the black-box setting, the attacker may make an unlimited number of queries to the malware detector without 
observing its internal operation.  
We permit access to detection confidence scores, which are returned alongside predictions even in the black-box 
setting. 
In the white-box setting, the attacker can additionally inspect the malware detector's source code. 
Such a strong assumption is needed for white-box attacks against neural network-based detectors that compute loss 
gradients with respect to the network's internal representation of the input file~\cite{kreuk2019deceiving,
lucas2021malware}.

\section{Experimental setup for malware detection case study} \label{app-sec:eval-setup}
In this appendix, we detail the experimental setup for our malware detection case study. 

\subsection{Datasets} \label{app-sec:eval-setup-datasets}
Though our methods are compatible with executable files of any format, in our experiments we focus on the 
\emph{Portable Executable (PE) format}~\cite{pespec}, since datasets, malware detection models and adversarial attacks 
are more extensively available for this format. 
Moreover, PE format is the standard for executables, object files and shared libraries in the Microsoft Windows operating system, making it an attractive target for malware authors. 
We use two PE datasets which are summarized in Table~\ref{tbl:data-summary} and described below.

\paragraph{\sleipnir} This dataset attempts to replicate data used in past work~\cite{aldujaili2018adversarial}, 
which was not published with raw samples. %
We were able to obtain the raw malicious samples from a public malware repository called VirusShare~\cite{virusshare} 
using the provided hashes.
However, since there is no similar public repository for benign samples, we followed established 
protocols~\cite{schultz2001data, kolter2006learning, kreuk2019deceiving} to collect a new set of benign samples. 
Specifically, we set up a Windows~7 virtual machine with over 300~packages installed using Chocolatey 
package manager~\cite{chocolatey}. 
We then extracted PE files from the virtual machine, which were assumed benign,\footnote{Chocolatey packages are validated against VirusTotal \cite{chocolatey2022security}.} and subsampled them to match 
the number of malicious samples. 
The dataset is randomly split into training, validation and test sets with a ratio of 60\%, 20\% and 20\% 
respectively.

\paragraph{\vtfeed} This dataset was first used in recent attacks on end-to-end ML-based malware 
detectors~\cite{lucas2021malware}.
It was collected from VirusTotal---a commercial threat intelligence service---by sampling PE files from the live 
feed over a period of two weeks in 2020. 
Labels for the files were derived from the 68 antivirus (AV) products aggregated on VirusTotal at the time of 
collection. 
Files were labeled \emph{malicious} if they were flagged malicious by 40 or more of the AV products, they were 
labeled \emph{benign} if they were not flagged malicious by any of the AV products, and any remaining files were 
excluded. 
Following \citeteam{lucas2021malware}, the dataset is randomly split into training, validation and test sets with a 
ratio of 80\%, 10\%, and 10\% respectively.

We note that \vtfeed\ comes with strict terms of use, which prohibit us from loading it on our high 
performance computing (HPC) cluster. 
As a result, we use \sleipnir\ for comprehensive experiments (e.g., varying $p_\del$, $\vec{\eta}$) on the HPC cluster, 
and \vtfeed\ for a smaller selection of experiments run on a local server.

\begin{table}
  \caption{Summary of datasets.}
  \label{tbl:data-summary}
  \begin{center}
    \small
    \begin{tabular}{llcccc}
      \toprule
                                &           & \multicolumn{3}{c}{Number of samples} \\
      \cmidrule(lr){3-5}
      Dataset                    & Label     & Train      & Validation & Test       \\
      \midrule
      \multirow{2}{*}{\sleipnir} & Benign    &    20\,948 &     7\,012 &     6\,999 \\
                                & Malicious &    20\,768 &     6\,892 &     6\,905 \\
      \midrule
      \multirow{2}{*}{\vtfeed}   & Benign    &   111\,258 &    13\,961 &    13\,926 \\
                                & Malicious &   111\,395 &    13\,870 &    13\,906 \\
      \bottomrule
    \end{tabular}
  \end{center}
\end{table}

\subsection{Malware detection models} \label{app-sec:eval-setup-models}
We experiment with malware detection models based on MalConv~\cite{raff2018malware}. 
MalConv was one of the first \emph{end-to-end} neural network models proposed for malware detection---i.e., 
it learns to classify directly from raw byte sequences, rather than relying on manually engineered features. 
Architecturally, it composes a learnable embedding layer with a shallow convolutional network. 
A large window size and stride of 500 bytes are employed to facilitate scaling to long byte sequences. 
Though MalConv is compatible with arbitrarily long byte sequences in principle, we truncate all inputs to 2MB to 
support training efficiency.
We use the original parameter settings and training procedure~\cite{raff2018malware}, except where specified in 
Appendix~\ref{app-sec:malconv-params}.

Using MalConv as a basis, we consider three models as described below.

\paragraph{\ns} A vanilla non-smoothed (\ns) MalConv model. 
This model serves as a non-certified, non-robust baseline---i.e., no specific techniques are employed to improve 
robustness to evasion attacks and certification is not supported. 

\paragraph{\rsabn} A smoothed MalConv model using the \emph{randomized ablation} smoothing mechanism proposed 
by~\citeteam{levine2020robustness} and reviewed in Appendix~\ref{app-sec:rs-abn}.
This model serves as a certified robust baseline, albeit covering a more restricted threat model than the edit distance 
threat model we propose in Section~\ref{sec:formulation}. 
Specifically, it supports robustness certification for the Hamming distance threat model, where the adversary is limited to substitution edits ($O = \{\sub\}$). 
Since Levine and Feizi's formulation is for images, several modifications are required to support malware detection 
as described in Appendix~\ref{app-sec:eval-comp-efficiency}. 
To improve convergence, we also apply gradient clipping when learning parameters in the embedding layer
(see Appendix~\ref{app-sec:eval-comp-efficiency}). 
We consider variants of this model for different values of the ablation probability $p_\ab$.

\paragraph{\rsdel} A smoothed MalConv model using our proposed randomized deletion smoothing mechanism. 
This model supports robustness certification for the generalized edit distance threat 
model where $O \subseteq \{\del, \ins, \sub\}$. 
We consider variants of this model for different values of the deletion probability $p_\del$, decision thresholds 
$\vec{\eta}$, and whether deletion\slash certification is performed at the byte-level (\textsc{Byte}) or chunk-level 
(\textsc{Insn}). 
We perform chunking as illustrated in Figure~\ref{fig:insn-del-exp}---i.e., we chunk bytes that correspond to distinct 
machine instructions using the Ghidra disassembler.

\subsection{Controlling false positive rates}
Malware detectors are typically tuned to achieve a low false positive rate (FPR) (e.g., less than $0.1\textrm{--}1\%$)
since producing too many false alarms is a nuisance to users.%
\footnote{\url{https://www.av-comparatives.org/testmethod/false-alarm-tests/}}
To make all malware detection models comparable, we
calibrate the FPR to 0.5\% on the test set for the experiments reported in Appendix~\ref{app-sec:eval-certificate} and 0.5\% on the validation set for the experiments reported in Appendix~\ref{app-sec:eval-attack} unless 
otherwise noted. %
This calibration is done by adjusting the decision threshold of the base MalConv model.

\subsection{Compute resources}
Experiments for the \sleipnir\ dataset were run a high performance computing (HPC) cluster, where the requested  
resources varied depending on the experiment. 
We generally requested a single NVIDIA P100 GPU when training and certifying models. 
Experiments for the \vtfeed\ dataset were run on a local server due to restrictive terms of use. 
Compute resources and approximate wall clock running times
are reported in Tables~\ref{tbl:compute-train} and~\ref{tbl:compute-certification} for training and certification  
for selected parameter settings. 
Running times for other parameters settings are lower than the ones reported in these tables.

\begin{table}
  \caption{
    Compute resources used for training. 
    Note that the wall clock times reported here are for an unoptimized implementation where the smoothing mechanism 
    is executed on the CPU.
  }
  \label{tbl:compute-train}
  \begin{center}
    \small
    \begin{tabular}{lm{0.25\linewidth}lm{0.11\linewidth}lm{0.22\linewidth}}
      \toprule
      Dataset                    & Requested resources
                                 & Model                                                                                            & Parameters                           & Time   & Notes                                          \\
      \midrule
      \multirow{8}{*}{\sleipnir} & \multirow{8}{\linewidth}{1 NVIDIA P100 GPU, 4 cores on Intel Xeon Gold 6326 CPU}
                                 & \ns                                                                                              & --                                   & 22 hr  & Trained for 50 epochs, converged in 15 epochs  \\
      \cmidrule(lr){3-6}
                                 &
                                 & \rsdel                                                                                           & \textsc{Byte},\newline $p_\del=95\%$ & 39 hr  & Trained for 100 epochs, converged in 50 epochs \\
      \cmidrule(lr){3-6}
                                 &
                                 & \rsdel                                                                                           & \textsc{Insn},\newline $p_\del=95\%$ & 48 hr  & Trained for 100 epochs, converged in 20 epochs \\
      \cmidrule(lr){3-6}
                                 &
                                 & \rsabn                                                                                           & $p_\ab=95\%$                         & 40 hr  & Trained for 100 epochs, converged in 90 epochs \\
      \midrule
      \multirow{4}{*}{\vtfeed}   & \multirow{4}{\linewidth}{1 NVIDIA RTX3090 GPU, 6 cores on AMD Ryzen Threadripper PRO 3975WX CPU}
                                 & \ns                                                                                              & --                                   & 139 hr & Trained for 100 epochs, converged in 25 epochs \\
      \cmidrule(lr){3-6}
                                 &
                                 & \rsdel                                                                                           & \textsc{Byte},\newline $p_\del=97\%$ & 152 hr & Trained for 100 epochs, converged in 20 epochs \\
      \bottomrule
    \end{tabular}
  \end{center}
\end{table}

\begin{table}
  \caption{
    Compute resources used for certification on the test set.  
    The evaluation dataset is partitioned and processed on multiple compute nodes with the same specifications. 
    The reported time is the sum of wall times on each compute node.
    Note that the times reported are for an unoptimized implementation where the smoothing mechanism is executed on 
    the CPU.
  }
  \label{tbl:compute-certification}
  \begin{center}
    \small
    \begin{tabular}{lm{0.3\linewidth}lm{0.11\linewidth}l}
      \toprule
      Dataset                      & Requested resources
                                   & Model                                                                                              & Parameters                           & Time   \\
      \midrule
      \multirow{7.5}{*}{\sleipnir} & \multirow{7.5}{\linewidth}{1 NVIDIA P100 GPU, 12 cores on Intel Xeon Gold 6326 CPU}
                                   & \ns                                                                                                & --                                   & 5 min  \\
      \cmidrule(lr){3-5}
                                   &
                                   & \rsdel                                                                                             & \textsc{Byte},\newline $p_\del=95\%$ & 65 hr  \\
      \cmidrule(lr){3-5}
                                   &
                                   & \rsdel                                                                                             & \textsc{Insn},\newline $p_\del=95\%$ & 140 hr \\
      \cmidrule(lr){3-5}
                                   &
                                   & \rsabn                                                                                             & $p_\ab=95\%$                         & 210 hr \\
      \midrule
      \multirow{3.5}{*}{\vtfeed}   & \multirow{3.5}{\linewidth}{1 NVIDIA RTX3090 GPU, 6 cores on AMD Ryzen Threadripper PRO 3975WX CPU}
                                   & \ns                                                                                                & --                                   & 4 min  \\
      \cmidrule(lr){3-5}
                                   &
                                   & \rsdel                                                                                             & \textsc{Byte},\newline $p_\del=97\%$ & 500 hr \\
      \bottomrule
    \end{tabular}
  \end{center}
\end{table}

\subsection{Parameter settings} \label{app-sec:malconv-params}

We specify the parameter settings and training procedure for MalConv, which is used standalone in \ns, and as a base 
model for the smoothed models \rsdel\ and \rsabn. 
Table~\ref{tbl:malconv-parameters} summarizes our setup, which is consistent across all three models except 
where specified. 
We follow the authors of MalConv \cite{raff2018malware} when setting parameters for the model and the optimizer, 
however we set a larger maximum input size of 2MiB to accommodate larger inputs without clipping.
Due to differences in available GPU memory for the \sleipnir\ and \vtfeed\ experiments, we use a larger batch 
size for \vtfeed\ than for \sleipnir. 
We also set a higher limit on the maximum number of epochs for \vtfeed, as it is a larger dataset, although the 
\ns\ and \rsdel\ models converge within 50~epochs for both datasets. 
To stabilize training for the smoothed models (\rsdel\ and \rsabn), we modify the smoothing mechanisms during 
\emph{training only} to ensure at least 500 raw bytes are preserved. 
This may limit the number of deletions for \rsdel\ and the number of ablated (masked) bytes for \rsabn. 
For \rsabn, we clip the gradients for the embedding layer to improve convergence (see 
Appendix~\ref{app-sec:eval-comp-efficiency}).

\begin{table}
  \caption{
    Parameter settings for MalConv, the optimizer and training procedure.
    Parameter settings are consistent across all malware detection models (\ns, \rsdel, \rsabn) except where
    specified.
  }
  \label{tbl:malconv-parameters}
  \begin{center}
  \small
  \begin{tabular}{cll}
    \toprule
                                        & Parameter          & Values                                     \\\midrule
    \multirow{5}{*}{MalConv}   & Max input size              & 2097152                                             \\\cmidrule(lr){2-3}
                                        & Embedding size              & 8                                                   \\\cmidrule(lr){2-3}
                                        & Window size                 & 500                                                 \\\cmidrule(lr){2-3}
                                        & Channels                    & 128                                                 \\ \midrule
    \multirow{5}{*}{Optimizer} & Python class                & \texttt{torch.optim.SGD}                            \\ \cmidrule(lr){2-3}
                                        & Learning rate               & 0.01                                                \\\cmidrule(lr){2-3}
                                        & Momentum                    & 0.9                                                 \\\cmidrule(lr){2-3}
                                        & Weight decay                & 0.001                                               \\\midrule
    \multirow{7}{*}{Training}  & Batch size                  & 24 (\sleipnir), 32 (\vtfeed)                        \\ \cmidrule(lr){2-3}
                                        & Max.\ epoch                 & 50 (\sleipnir), 100 (\vtfeed)                       \\\cmidrule(lr){2-3}
                                        & Min.\ preserved bytes       & 500 (\rsdel, \rsabn), NA (\ns)                      \\\cmidrule(lr){2-3}
                                        & Embedding gradient clipping & 0.5 (\rsabn), $\infty$ (\rsdel, \ns)                \\\cmidrule(lr){2-3}
                                        & Early stopping              & If validation loss does not improve after 10 epochs \\\bottomrule
  \end{tabular}
  \end{center}
\end{table}

\section{Evaluation of robustness certificates for malware detection} \label{app-sec:eval-certificate}

In this appendix, we evaluate the robustness guarantees and accuracy of \rsdel for malware detection. 
We consider two instantiations of the edit distance threat model. 
First, in Appendix~\ref{app-sec:eval-cert-lev-dist}, we consider the Levenshtein distance threat model, where the 
attacker's elementary edits are unconstrained and may include deletions, insertions and substitutions. 
Then, in Appendix~\ref{app-sec:eval-cert-hamm-dist}, we consider the more restricted Hamming distance threat model, where an 
attacker is only able to perform substitutions. 
We summarize our findings in Appendix~\ref{app-sec:eval-cert-summ}. Overall, we find that \rsdel\ 
generates robust predictions with minimal impact on model accuracy for the Levenshtein distance threat model, 
and outperforms \rsabn\ \cite{levine2020robustness} for the Hamming distance threat model.

We report the following quantities in our evaluation:
\begin{itemize}
  \item \emph{Certified radius (CR).} The radius of the largest robustness certificate that can be issued for a given 
  input, model and certification method. 
  Note that this is a conservative measure of robustness since it is \emph{tied to the certification method}. 
  The \emph{median CR} is reported on the test set. 
  \item \emph{Certified accuracy} \cite{lecuyer2019certified, cohen2019certified}, also known as 
  \emph{verified-robust accuracy} \cite{wong2018scaling, leino2021globally}, evaluates robustness certificates and 
  accuracy of a model simultaneously with respect to a test set. 
  It is defined as the fraction of instances in the test set $\mathbb{D}$ for which the model $f$'s 
  prediction is correct \emph{and} certified robust at radius~$\radius$ or greater:
  \begin{align}
    \textsc{CertAcc}_r(\mathbb{D}) = \sum_{(\vec{x}, y) \in
      \mathbb{D}}\frac{\ind{f(\vec{x}) = y}\ind{\mathrm{CR}(\vec{x}) \geq r}}{\abs{\mathbb{D}}} \label{eqn:vra}
  \end{align}
  where $\mathrm{CR}(\vec{x})$ denotes the certified radius for input $\vec{x}$ returned by the certification method. 
  \item \emph{Clean accuracy.} The fraction of instances in the test set for which the model's prediction 
  is correct. 
\end{itemize}

We briefly mention default parameter settings for the experiments presented in this appendix. 
When approximating the smoothed models (\rsdel\ and \rsabn) we sample $n_\mathrm{pred} = 1000$ perturbed
inputs for prediction and $n_\mathrm{bnd} = 4000$ perturbed inputs for certification, while setting the 
significance level $\alpha$ to $0.05$. 
Unless otherwise specified, we set the decision thresholds for the smoothed models so that $\vec{\eta} = 0$. 
After fixing $\vec{\eta}$, the decision thresholds for the base models are tuned to yield a false positive rate of 
0.5\%. 
We note that the entire test set is used when reporting metrics and summary statistics in this appendix. 

\subsection{Levenshtein distance threat model} \label{app-sec:eval-cert-lev-dist}
We first present results for the Levenshtein distance threat model, where the attacker's elementary edits are 
unconstrained  ($O = \{\del, \ins, \sub\}$). 
We vary three parameters associated with \rsdel: the deletion probability~$p_\del$, the decision 
thresholds of the smoothed model $\vec{\eta}$, and the level of sequence chunking (i.e., whether sequences are 
chunked at the byte-level or instruction-level). 
We use \ns\ as a baseline as there are no prior certified defenses for the Levenshtein distance threat model to our 
knowledge.

\begin{figure}
  \begin{minipage}[t]{0.49\textwidth}
    \centering
    \includegraphics[width=0.92\linewidth]{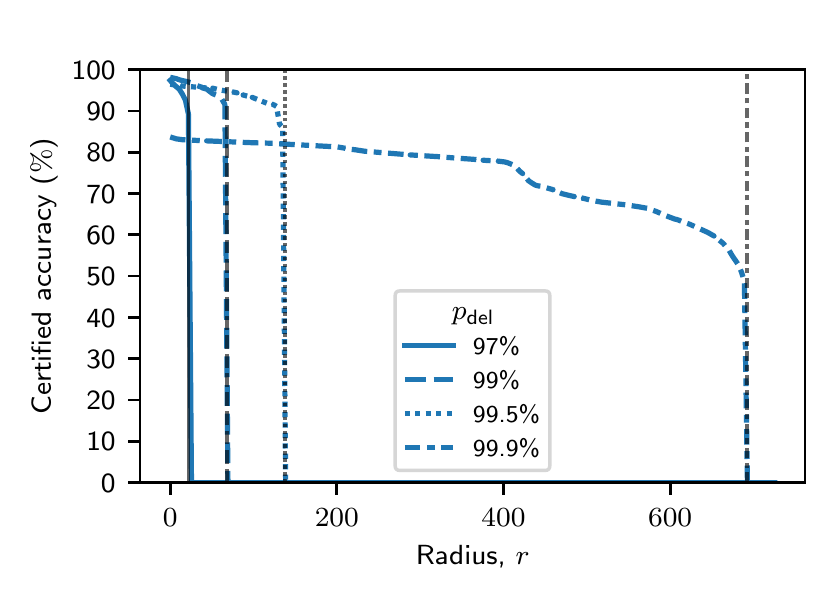}
  \end{minipage}
  \hfill
  \begin{minipage}[t]{0.49\textwidth}
    \centering
    \includegraphics[width=0.92\linewidth]{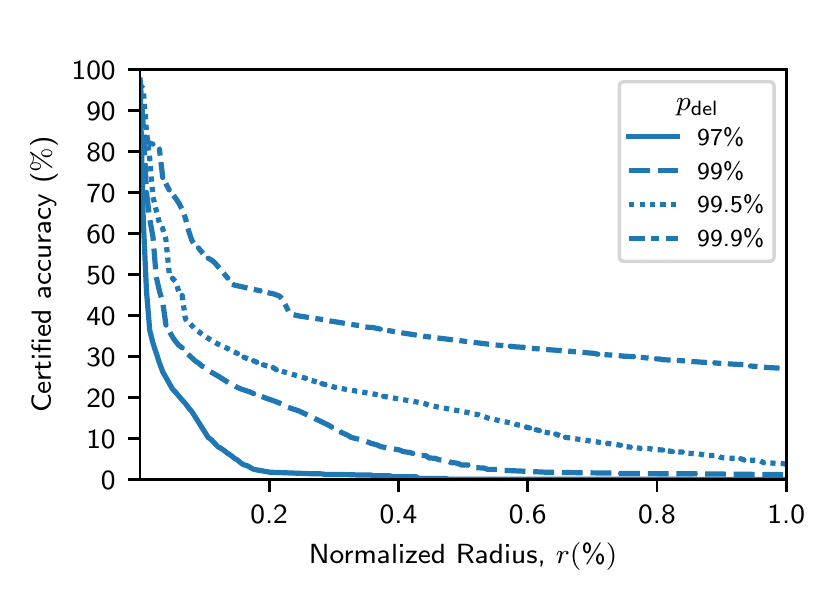}
  \end{minipage}
  \caption{%
    Certified accuracy for \rsdel\ as a function of the radius in bytes (left horizontal axis), radius normalized by 
    file size (right horizontal axis) and byte deletion probability $p_\del$ (line styles).
    The results are plotted for the \sleipnir\ test set under the byte-level Levenshtein distance threat model 
    (with $O = \{\del, \ins, \sub\}$) .  
    The grey vertical lines in the left plot represent the best achievable certified radius for \rsdel\ (setting 
    $\mu_y = 1$ in the expressions in Table~1).
  }
  \label{fig:ca-lev-sleipnir-pdel}
\end{figure}

\begin{figure}
  \begin{minipage}[t]{0.49\textwidth}
    \centering
    \includegraphics[width=0.92\linewidth]{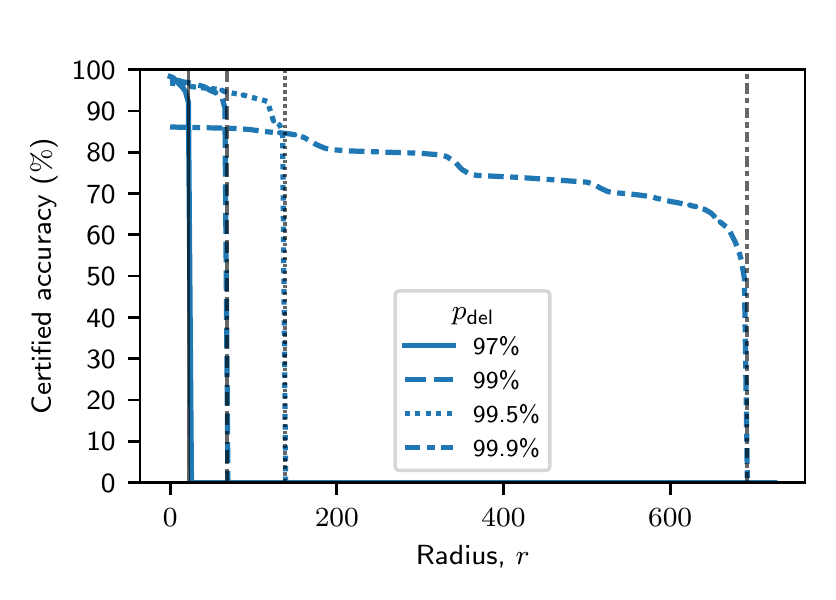}
  \end{minipage}
  \hfill
  \begin{minipage}[t]{0.49\textwidth}
    \centering
    \includegraphics[width=0.92\linewidth]{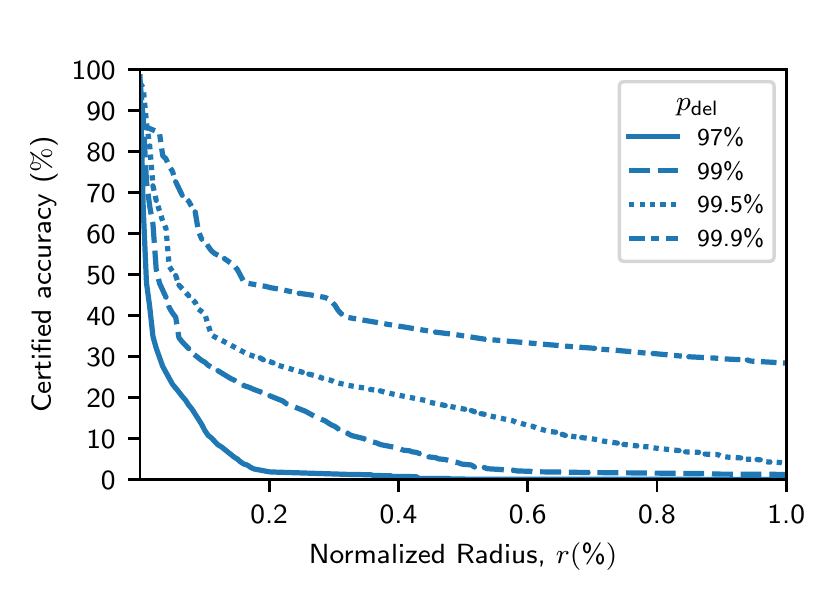}
  \end{minipage}
  \caption{%
    Certified accuracy for \rsdel\ with chunk-level deletion (\textsc{Insn}) as a function of the radius in chunks 
    (left horizontal axis), radius normalized by sequence length in chunks (right horizontal axis) and chunk deletion 
    probability $p_\del$ (line styles).
    The results are plotted for the \sleipnir\ test set under the chunk-level Levenshtein distance threat 
    model (with $O = \{\del, \ins, \sub\}$).
    The grey vertical lines in the left plot represent the best achievable certified radius for \rsdel\ (setting 
    $\mu_y = 1$ in the expressions in Table~1).
  }
  \label{fig:ca-lev-sleipnir-pdel-insn}
\end{figure}

\paragraph{Certified accuracy}
Figure~\ref{fig:ca-lev-sleipnir-pdel} plots the certified accuracy of \rsdel\ using byte-level deletion as a function 
of the radius %
(left horizontal axis), radius normalized by file size (right horizontal axis)
on the \sleipnir\ dataset for several values of $p_\del$. 
We observe that the curves for larger values of $p_\del$ approximately dominate the curves for smaller values of 
$p_\del$, for $p_\del \leq 99.5\%$ (i.e., the accuracy is higher or close for all radii).
This suggests that the robustness of \rsdel\ can be improved without sacrificing accuracy by increasing $p_\del$ up to 
99.5\%. 
However, for the larger value $p_\del = 99.9\%$, we observe a drop in certified accuracy of around 10\% for smaller 
radii and an increase for larger radii. %
By normalizing with respect to the file size, we can see that our certificate is able to certify up to $1\%$ of the file size.
We also include an analogous plot for chunk-level deletion in Figure~\ref{fig:ca-lev-sleipnir-pdel-insn} 
which demonstrates similar behavior. 
We note that chunk-level deletion arguably provides stronger guarantees, since the effective 
radius for chunk-level Levenshtein distance is larger than for byte-level Levenshtein distance.

It is interesting to relate these certification results to published evasion attacks. 
Figure~\ref{fig:ca-lev-sleipnir-pdel} shows that we can achieve a certified accuracy in excess of 90\% at a 
Levenshtein distance radius of 128 bytes when $p_\del = 99.5\%$. 
This radius is larger than the median Levenshtein distance of two attacks that manipulate headers of PE files  
\cite{demetrio2019explaining,nisi2021lost} (see Table~\ref{tbl:adv-attacks} in Appendix~\ref{app-sec:eval-attack}). 
We can therefore provide reasonable robustness guarantees against these two attacks. 
However, a radius of 128 bytes is orders of magnitude smaller than the median Levenshtein distances of other published 
attacks which range from tens of KB~\cite{lucas2021malware,kreuk2019deceiving} to several 
MB~\cite{demetrio2021functionality} (also reported in Table~\ref{tbl:adv-attacks}). 
While some of these attacks arguably fall outside an edit distance constrained threat model, we consider them in our 
empirical evaluation of robustness in Appendix~\ref{app-sec:eval-attack}.

\paragraph{Clean accuracy and abstention rates}
Table~\ref{tbl:cr-statistics} reports clean accuracy for \rsdel\ and the non-certified \ns\ baseline. 
It also reports abstention rates for \rsdel, the median certified radius (CR), and the median 
certified radius normalized by file size (NCR). 
We find that clean accuracy for \sleipnir\ follows similar trends as certified accuracy: it is relatively 
stable for $p_\del$ in the range 90--99.5\%, but drops by more than 10\% at $p_\del = 99.9\%$. 
We note that the clean accuracy of \rsdel\ (excluding $p_\del = 99.9\%$) is at most 3\% lower than 
the \ns\ baseline for \sleipnir\ and at most 7\% lower than the \ns\ baseline for \vtfeed. 
We observe minimal differences in the results for chunk-level (\textsc{Insn}) and byte-level (\textsc{Byte}) 
deletion smoothing, but note that the effective CR is larger for chunk-level smoothing, since each chunk may contain 
several bytes.

\begin{table}
  \centering
  \caption{Clean accuracy and robustness metrics for \rsdel\ as a function of the dataset (\sleipnir\ and \vtfeed), 
    deletion probability $p_\del$ and deletion level (\textsc{Byte} or \textsc{Insn}). 
    All metrics are computed on the test set. 
    Here ``abstain rate'' refers to the fraction of test instances for which \rsdel\ abstains (line~7 in 
    Figure~\ref{alg:estimation-algorithm}), and ``UB'' refers to an upper bound on the median CR for a 
    best case smoothed model (based on Table~\ref{tbl:certificate-O} with $\mu_y = 1$).
    A good tradeoff is achieved when $p_\del = 99.5\%$ for both the byte-level (\textsc{Byte}) and chunk-level 
    (\textsc{Insn}) certificates (highlighted in bold face below). 
  }
  \label{tbl:cr-statistics}
  \begin{center}
    \small
    \begin{tabular}{
        lll
        d{5.1}
        @{} d{2.2}
        d{3.0}
        @{} d{5.0}
        d{1.4}
      }\toprule
               &                          &                                    & \multicolumn{2}{c}{Clean accuracy}    & \multicolumn{2}{c}{Median CR}     & \mc{Median}                                          \\
      Dataset  & Model                    & \multicolumn{1}{l}{Parameters}    & \multicolumn{2}{c}{(Abstain rate) \%} & \multicolumn{2}{c}{(UB)} & \mc{NCR \%}                                                      \\
      \midrule
      \multirow{13}{*}[-2pt]{\sleipnir} 
      & \ns                      & $-$                                & 98.9                                  & \mc{$-$}                       & \multicolumn{1}{r}{$-$} & \mc{$-$}        & \mc{$-$}         \\
      \cmidrule{2-8}
      & \multirow{12}{*}[-2pt]{\rsdel} & \textsc{Byte}, $p_\del = 90\%$     & 97.1                                  & (0.2)                          & 6                       & (6)             & 0.0023           \\
      &                          & \textsc{Byte}, $p_\del = 95\%$     & 97.8                                  & (0.0)                          & 13                      & (13)            & 0.0052           \\
      &                          & \textsc{Byte}, $p_\del = 97\%$     & 97.4                                  & (0.1)                          & 22                      & (22)            & 0.0093           \\
      &                          & \textsc{Byte}, $p_\del = 99\%$     & 98.1                                  & (0.1)                          & 68                      & (68)            & 0.0262           \\
      &                          & \textsc{Byte}, $p_\del = 99.5\%$   & \boldcell 96.5                        & \boldcell (0.2)                & \boldcell 137           & \boldcell (138) & \boldcell 0.0555 \\
      &                          & \textsc{Byte}, $p_\del = 99.9\%$   & 83.7                                  & (3.4)                          & 688                     & (692)           & 0.2269           \\
      \cmidrule{3-8}
      &                          & \textsc{Insn}, $p_\del = 90\%$     & 97.9                                  & (0.1)                          & 6                       & (6)             & 0.0026           \\
      &                          & \textsc{Insn}, $p_\del = 95\%$     & 97.8                                  & (0.1)                          & 13                      & (13)            & 0.0056           \\
      &                          & \textsc{Insn}, $p_\del = 97\%$     & 98.3                                  & (0.0)                          & 22                      & (22)            & 0.0095           \\
      &                          & \textsc{Insn}, $p_\del = 99\%$     & 97.6                                  & (0.1)                          & 68                      & (68)            & 0.0292           \\
      &                          & \textsc{Insn}, $p_\del = 99.5\%$   & \boldcell 96.8                        & \boldcell (0.2)                & \boldcell 137           & \boldcell (138) & \boldcell 0.0589 \\
      &                          & \textsc{Insn}, $p_\del = 99.9\%$   & 86.1                                  & (0.2)                          & 689                     & (692)           & 0.2982           \\
      \midrule
      \multirow{3}{*}[-2pt]{\vtfeed} 
      & \ns                      & $-$                                & 98.9                                  & \mc{$-$}                       & \mc{$-$}                & \mc{$-$}        & \mc{$-$}         \\
      \cmidrule{2-8}
      & \multirow{2}{*}{\rsdel}  & \textsc{Byte}, $p_\del = 97\%$     & 92.1                                  & (0.9)                          & 22                      & (22)            & 0.0045            \\
      &                          & \textsc{Byte}, $p_\del = 99\%$     & 86.9                                  & (0.8)                          & 68                      & (68)            & 0.0122            \\
      \bottomrule
    \end{tabular}
  \end{center}
\end{table}

\paragraph{Accuracy under high deletion}
It may be surprising that \rsdel\ can maintain high accuracy even when deletion is aggressive.
We offer some possible explanations. 
First, we note that even with a high deletion probability of $p_\del = 99.9\%$, the smoothed model accesses almost 
all of the file in expectation, as it aggregates $n_\mathrm{pred} = 1000$ predictions from the base model
each of which accesses a random $0.1\%$ of the file in expectation. 
Second, we posit that malware detection may be ``easy'' for \rsdel\ on these datasets. 
This could be due to the presence of signals that are robust to deletion (e.g., file size or byte frequencies) or 
redundancy of signals (i.e., if a signal is deleted in one place it may be seen elsewhere).

\paragraph{Decision threshold}
We demonstrate how the decision thresholds $\vec{\eta}$ introduced in Section~\ref{sec:rs} can be used to trade off 
certification guarantees between classes. 
We consider normalized decision thresholds where $\sum_{y} \eta_y = 1$ and $\eta_y \in [0, 1]$. 
We only specify the value of $\eta_1$ when discussing our results, noting that $\eta_0 = 1 - \eta_1$ in our two-class 
setting.

Table~\ref{tbl:cr-statistics-smoothprob} provides error rates and robustness metrics for several values of $\eta_1$, 
using byte-level Levenshtein distance with $p_\del = 99.5\%$.
When varying $\eta_1$, we also vary the decision threshold of the base model to achieve a target 
false positive rate (FPR) of 0.5\%.%
Looking at the table, we see that $\eta_1$ has minimal impact on the false negative rate (FNR), which is  
stable around 7\%. 
However, there is a significant impact on the median CR (and theoretical upper bound), as reported separately for 
each class. 
The median CR is balanced for both the malicious and benign class when $\eta_1 = 50\%$, but favours the malicious 
class as $\eta_1$ is decreased. 
For instance when $\eta_1 = 5\%$ a significantly larger median CR is possible for malicious files (137 to 578) 
at the expense of the median CR for benign files (137 to 10). 
This asymmetry in the class-specific CR is a feature of the theory---that is, in addition to controlling a 
tradeoff between error rates of each class, $\eta_1$ also controls a tradeoff between the CR for 
each class (see Table~\ref{tbl:certificate-O}).

\begin{table}
  \centering
  \caption{%
    Impact of the smoothed decision threshold $\eta_1$ on false negative error rate (FNR) and median certified 
    radius (CR) for malicious and benign files. 
    The false positive rate (FPR) is set to a target value of 0.5\% by varying the decision threshold of the base 
    model. 
    The results are reported for \sleipnir\ with $p_\del = 99.5\%$ using byte-level Levenshtein distance. 
    ``UB'' refers to an upper bound on the median CR for a best case smoothed model (based on 
    Table~\ref{tbl:certificate-O} with $\mu_y = 1$).
  }
  \label{tbl:cr-statistics-smoothprob}
  \begin{center}
    \small
    \begin{tabular}{
        d{2.1}
        d{1.1}
        d{3.1}
        d{3.0}
        @{} d{6.0}
        d{3.0}
        @{} d{5.0}
      }\toprule
       & &  & \multicolumn{4}{c}{Median CR (UB)}  \\
      \cmidrule(lr){4-7}
      \mc{$\eta_1$ (\%)} & \mc{FNR (\%)} & \mc{FPR (\%)} & \multicolumn{2}{c}{Malicious} &  \multicolumn{2}{c}{Benign} \\
      \midrule
      50   &  6.8 &   0.5 &  137 &  (138) & 137 & (138)  \\
      25   &  6.9 &   0.5 &  275 &  (276) &  57 &  (57) \\
      10   &  6.8 &   0.5 &  455 &  (459) &  20 &  (21) \\
       5   &  6.6 &   0.5 &  578 &  (597) &  10 &  (10)  \\
       1   &  7.1 &   0.5 &  582 &  (918) &   1 &   (2) \\
       0.5 &  6.9 &   0.5 &  506 & (1057) &   0 &   (0) \\
      \bottomrule
    \end{tabular}
  \end{center}
\end{table}

\begin{figure}
  \centering
  \includegraphics[width=0.65\linewidth]{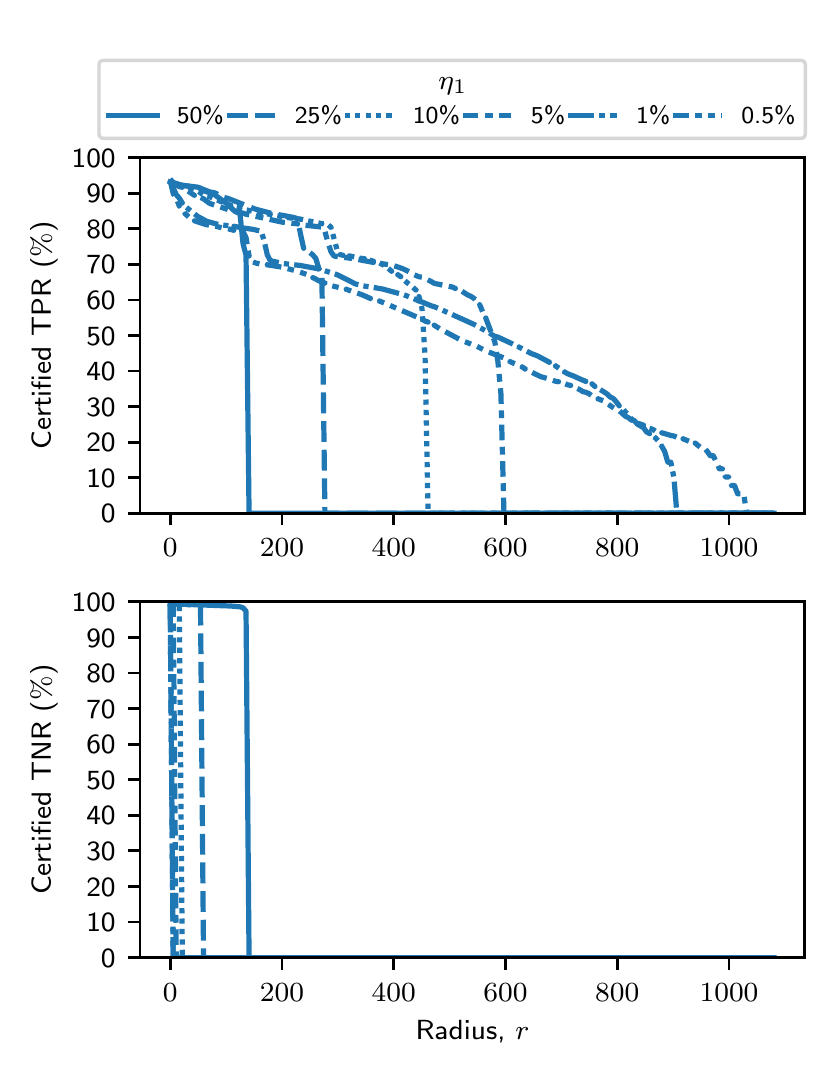}
  \caption{
    Certified true positive rate (TPR) and true negative rate (TNR) of \rsdel\ as a function of the certificate 
    radius $\radius$ (horizontal axis) and the decision threshold $\eta_1$ (line style). 
    The results are plotted for the \sleipnir\ test set for byte-level deletion (\textsc{Byte}) with $p_\del = 99.5\%$ 
    under the Levenshtein distance threat model (with $O = \{\del, \ins, \sub\}$).
    It is apparent that $\eta_1$ controls a tradeoff in the certified radius between the malicious (measured 
    by TPR) and benign (measured by TNR) classes.
    Note that in this setting, a non-smoothed, non-certified model (\ns) achieves a clean TPR and TNR of 98.2\% 
    and 99.5\% respectively.
  }
  \label{fig:ctprtnr-lev-sleipnir-eta-byte}
\end{figure}

Figure~\ref{fig:ctprtnr-lev-sleipnir-eta-byte} plots the certified true positive rate (TPR) and true negative rate 
(TNR) of \rsdel\ on the \sleipnir\ dataset for several values of $\eta_1$. 
The certified TPR and TNR can be interpreted as class-specific analogues of the certified accuracy. 
Concretely, the certified TPR (TNR) at radius $\radius$ is the fraction of malicious (benign) instances in the test 
set for which the model's prediction is correct \emph{and} certified robust at radius $\radius$. 
The certified TPR and TNR jointly measure accuracy and robustness and complement the metrics reported in 
Table~\ref{tbl:cr-statistics-smoothprob}. 
Looking at Figure~\ref{fig:ctprtnr-lev-sleipnir-eta-byte}, we see that the certified TNR curves drop more rapidly to 
zero than the certified TPR curves as $\eta_1$ decreases. 
Again, this suggests decreasing $\eta_1$ sacrifices the certified radii of benign instances to increase the certified 
radii of malicious instances.
We note that the curves for $\eta_1 = 50\%$ correspond to the same setting as the certified accuracy curve 
in Figure~\ref{fig:ca-lev-sleipnir-pdel} (with $p_\del = 99.5\%$).

\subsection{Hamming distance threat model} \label{app-sec:eval-cert-hamm-dist}
We now turn to the more restricted Hamming distance threat model, where the attacker is limited to performing 
substitutions only ($O = \{\sub\}$).
We choose to evaluate this threat model as it is covered in previous work on randomized smoothing, called 
\emph{randomized ablation} \cite{levine2020robustness} (abbreviated \rsabn), and can serve as a baseline for comparison with our method. 
Recall that we adapt \rsabn\ for malware detection by introducing a parameter called $p_\ab$, which is the 
fraction of bytes that are ``ablated'' (replaced by a special masked value) (see Appendix~\ref{app-sec:eval-setup-models}).
This parameter is analogous to $p_\del$ in \rsdel, except that the number of ablated bytes is deterministic in \rsabn, 
whereas the number of deleted bytes is random in \rsdel. 
We compare \rsdel\ and \rsabn\ for varying values of $p_\del$ and $p_\ab$ using the \sleipnir\ dataset and 
byte-level Hamming distance.

\paragraph{Certified accuracy}
Figure~\ref{fig:ca-hamming-sleipnir-cmp} plots the certified accuracy of \rsdel\ and \rsabn\ for three values of 
$p_\del$ and $p_\ab$. 
We observe that the certified accuracy is uniformly larger for our proposed method \rsdel\ than for \rsabn\ 
when $p_\del = p_\ab$. 
The superior certification performance of \rsdel\ is somewhat surprising given it is not optimized for the Hamming 
distance threat model. 
One possible explanation relates to the learning difficulty of \rsabn\ compared with \rsdel. 
Specifically, we find that stochastic gradient descent is slower to converge for \rsabn\ despite our attempts to 
improve convergence (see Appendix~\ref{app-sec:eval-comp-efficiency}). 
Recall, that \rsdel\ provides certificates for any of the threat models in 
Table~\ref{tbl:certificate-O}---in addition to the Hamming distance certificate---without needing to modify the 
smoothing mechanism. 

\paragraph{Tightness}
\rsabn\ is provably tight, in the sense that it is not possible to issue a larger Hamming distance certificate 
unless more information is made available to the certification mechanism or the ablation smoothing mechanism is 
changed. 
This tightness result for \rsabn, together with the empirical results in Figure~\ref{fig:ca-hamming-sleipnir-cmp}, 
suggests that \rsdel\ produces certificates which are tight or close to tight in practice, at least for the Hamming 
distance threat model. 
This is an interesting observation, since it is unclear how to derive a tight, computationally tractable certificate for \rsdel.

\subsection{Summary} \label{app-sec:eval-cert-summ}

Our evaluation shows that \rsdel\ provides non-trivial robustness guarantees with a low impact on accuracy.
The certified radii we observe are close to the best radii theoretically achievable using our mechanism. 
For the Levenshtein byte-level edit distance threat model, we obtain radii of a few hundred bytes in size, 
which can certifiably defend against attacks that edit headers of PE files \cite{demetrio2019explaining,
nisi2021lost,demetrio2021adversarial}.
However, certifying robustness against more powerful attacks that modify thousands or millions of bytes 
remains an open challenge. 
By varying the detection threshold, we show that certification can be performed asymmetrically for benign and 
malicious instances.
This can boost the certified radii of malicious instances by a factor of 4 in some cases. 
While there are no prior methods to use as baselines for the Levenshtein distance threat model, our comparisons 
with \rsabn~\cite{levine2020robustness} for the Hamming distance threat model show that \rsdel\ outperforms 
\rsabn\ in terms of both accuracy and robustness.

\section{Evaluation of robustness to published attacks} \label{app-sec:eval-attack}

In this appendix, we empirically evaluate the robustness of \rsdel\ to several published evasion attacks. 
By doing so, we aim to provide a more complete picture of robustness, as our certificates are conservative and 
may \emph{underestimate} robustness to real attacks, which are subject to additional constraints (e.g., 
maintaining executability, preserving a malicious payload, etc.).
We introduce the attacks in Appendix~\ref{app-sec:eval-attack-attacks}, provide details of the experimental setup in 
Appendix~\ref{app-sec:eval-attack-setup} and discuss the results in Appendix~\ref{app-sec:eval-attack-results}.

\subsection{Attacks covered} \label{app-sec:eval-attack-attacks}

\begin{table}[ht]
\centering
\caption{
    Evasion attacks used in our evaluation. 
    The \emph{attack distance} refers to the median Levenshtein distance computed on a set of 500 attacked files 
    from the \sleipnir\ test set. 
    We use a closed source implementation of \disp{} and open source implementations of the 
    remaining attacks based on \texttt{secml-malware}~\cite{demetrio2021secml}. 
}
\label{tbl:adv-attacks}
\begin{center}
  \small
  \begin{tabular}{
      l
      l
      d{2.4}
      l
      L{5cm}
      } \toprule
      Attack
      & \bigcell{l}{Supported\\settings} & \mc{\bigcell{l}{Attack\\distance}} & Optimizer  & Description \\
      \midrule
      \disp{} \cite{lucas2021malware}
      & \bigcell{l}{White-box,\\black-box}  &                  17.2{\ \text{KB}} & Gradient-guided & 
      Disassembles the PE file and displaces chunks of code to a new section, replacing the original code with 
      semantic nops. \\
      \midrule
      \slack{} \cite{kreuk2019deceiving}
      & \bigcell{l}{White-box}  &                  34.7{\ \text{KB}} & \bigcell{l}{Fast Gradient Sign\\Method \cite{goodfellow2015explaining}} &
      Replaces non-functional bytes in slack regions or the overlay of the PE file with adversarially-crafted 
      noise. \\
      \midrule
      \hdos{} \cite{demetrio2019explaining}
      & \bigcell{l}{White-box,\\black-box}  &                   58.0{\ \text{B}} & Genetic algorithm & 
      Manipulates bytes in the DOS header of the PE file which are not used in modern Windows. \\
      \midrule
      \hfield{} \cite{nisi2021lost} 
      & \bigcell{l}{White-box,\\black-box}  &                   17.0{\ \text{B}} & Genetic algorithm & 
      Manipulates fields in the header of the PE file (debug information, section names, checksum, etc.) which 
      do not impact functionality. \\
      \midrule
      \secinj{} \cite{demetrio2021functionality} 
      & Black-box                           &                  2.10{\ \text{MB}} & Genetic algorithm & 
      Appends sections extracted from benign files to the end of a malicious PE file and modifies the header 
      accordingly. \\
      \bottomrule
  \end{tabular}
\end{center}
\end{table}

We consider five recently published attacks designed for evading static PE malware detectors as summarized in 
Table~\ref{tbl:adv-attacks}. 
The attacks cover a variety of edit distance magnitudes from tens of bytes to millions of bytes. 
While attacks that edit millions of bytes arguably fall outside our edit distance-constrained threat model, we 
include one such attack (\secinj) to test the limits of our methodology. 
We note that four of the five attacks are able to operate in a black-box setting and can therefore be applied 
directly to \rsdel. 
However, the white-box attacks are designed for neural network malware detectors with a specific architecture. 
In particular, they assume the network receives a raw byte sequence as input, that the initial layer is an 
embedding layer, and that gradients can be computed with respect to the output of the embedding layer. 
Although these architectural assumptions are satisfied by the base MalConv model, they are not satisfied by \rsdel, 
because additional operations are applied before the embedding layer and the aggregation of base model predictions 
is not differentiable. 
In Appendix~\ref{app-sec:adapt-wb}, we adapt the white-box attacks for \rsdel\ by applying two tricks: (1)~we apply 
the smoothing mechanism \emph{after} the embedding layer, and (2)~we replace majority voting with soft aggregation 
following \citeteam{salman2019provably}.

\subsection{Experimental setup} \label{app-sec:eval-attack-setup}
Since some of the attacks take hours to run for a single file, we use smaller evaluation sets containing malware 
subsampled from the test sets in Table~\ref{tbl:data-summary}. 
The evaluation set we use for \sleipnir\ consists of 500~files, and the one for \vtfeed\ consists of 100~files 
(matching~\cite{lucas2021malware}). 
We note that our evaluation sets are comparable in size to prior work~\cite{kolosnjaji2018adversarial,
kreuk2019deceiving,suciu2019exploring}. 
For each evaluation set, we report attack success rates against malware detectors trained on the same dataset. 

Since all attacks employ greedy optimization with randomization, they may fail on some runs, but succeed on 
others.
We therefore repeat each attack 5 times per file and use the best performing attacked file in our evaluation.  
We define the attack success rate as the proportion of files initially detected as malicious for which at least one 
of the 5 attack repeats is successful at evading detection. 
Lower attack success rates correlate with improved robustness against attacks. 
We permit all attacks to run for up to 200 attack iterations of the internal optimizer. 
Early stopping is enabled for those attacks that support it (\disp, \slack, \secinj), which means the attack 
terminates as soon as the model's prediction flips from malicious to benign.

Where possible, we run \emph{direct attacks} against \rsdel\ and compare success rates against \ns\ as a baseline. 
We also consider \emph{transfer attacks} from \ns\ to \rsdel\ as an important variation to the threat model, where 
an attacker has limited access to the target \rsdel\ during attack optimization. 
When running direct attacks against \rsdel, we use a reduced number of Monte Carlo samples ($n_\mathrm{pred} = 100$) to 
make the computational cost of the attacks more manageable. 
For both direct and transfer attacks against \rsdel, we set $p_\del = 97\%$ and perform deletion and certification at 
the byte-level (\textsc{Byte}).

\subsection{Results} \label{app-sec:eval-attack-results}
The results for direct attacks against \rsdel\ are presented in Table~\ref{tbl:direct-attack-summary}. 
For both the \sleipnir\ and \vtfeed\ datasets, we observe that the robustness of \rsdel\ is superior (or equal) to 
\ns\ against four of the six attacks. 
The two cases where \rsdel's robustness drops compared to \ns\ are for the strongest attacks: \slack\ and \secinj. 
The results for transfer attacks from \ns\ to \rsdel\ are presented in Table~\ref{tbl:transfer-attack-summary}. 
Almost all of the attacks transfer poorly to \rsdel. 
In most cases the attack success rates drop to zero or single digit percentages.
We hypothesize that \slack\ and \secinj\ make such drastic changes to the original binary that they can overwhelm the 
malicious signal---enough to cross the decision boundary---akin to a good word attack~\cite{lowd2005good}. %
We find that \hdos\ and \hfield\ are ineffective for both \rsdel\ and the baseline \ns. Both attacks change up to 
58~bytes in the header, and tend to fall within our certifications.

\begin{table}
  \caption{
    Success rates of direct attacks against \rsdel\ and the \ns\ baseline. 
    A lower success rate is better from our perspective as a defender, as it means the model is more robust to the 
    attack. 
    The model that achieves the lowest success rate for each attack\slash dataset is highlighted in boldface.
  }
  \label{tbl:direct-attack-summary}
  \begin{center}
  \small
  \begin{tabular}{lll*{2}{d{3.3}}}
    \toprule
      & 
        &           & \multicolumn{2}{c}{Attack success rate (\%)}\\
    \cmidrule{4-5}
    Setting  
      & Attack 
        & Dataset   & \mc{\ns} & \mc{\rsdel} \\
    \midrule
    \multirow{4.4}{*}{White-box} 
      & \multirow{2}{*}{\disp{}~\cite{lucas2021malware}}
        & \sleipnir & 73.8   &   \boldcell 56.7 \\
      & 
        & \vtfeed   & 94.1   &   \boldcell 74.5   \\
    \cmidrule{2-5}
      & \multirow{2}{*}{\slack{}~\cite{kreuk2019deceiving}} 
        & \sleipnir & \boldcell 57.9   &   85.3  \\
      & 
        & \vtfeed   & 96.0   &   \boldcell 43.9  \\
    \midrule
    \multirow{9.2}{*}{Black-box} 
      & \multirow{2}{*}{\hdos{}~\cite{demetrio2019explaining}}
        & \sleipnir &    0.0   &     0.0   \\
      &
        & \vtfeed   &    0.0   &     0.0   \\
    \cmidrule{2-5}
      & \multirow{2}{*}{\hfield{}~\cite{nisi2021lost}} 
        & \sleipnir &    0.607 &     \boldcell 0.0   \\
      &
        & \vtfeed   &    0.990 &     \boldcell 0.0   \\
    \cmidrule{2-5}
      & \multirow{2}{*}{\disp{}~\cite{lucas2021malware}} 
        & \sleipnir &    0.809 &     \boldcell 0.0   \\
      &
        & \vtfeed   &   10.9   &     \boldcell 0.0   \\
    \cmidrule{2-5}
      & \multirow{2}{*}{\secinj{}~\cite{demetrio2021functionality}}
        & \sleipnir &   99.2   &    \boldcell 54.1   \\
      &
        & \vtfeed   &   \boldcell 76.2   &   100.0   \\
    \bottomrule
  \end{tabular}
  \end{center}
\end{table}

\begin{table}
  \caption{Success rates of attacks transferred from \ns\ to \rsdel.}
  \label{tbl:transfer-attack-summary}
  \begin{center}
  \small
  \begin{tabular}{lll*{2}{d{3.3}}}
    \toprule
      & 
        &           & \multicolumn{2}{c}{Attack success rate (\%)}\\
    \cmidrule{4-5}
    Setting  
      & Attack 
        & Dataset   & \mc{\ns} & \mc{\rsdel} \\
    \midrule
    \multirow{4.4}{*}{White-box} 
      & \multirow{2}{*}{\disp{}~\cite{lucas2021malware}}
        & \sleipnir & 73.8   &   0.414 \\
      & 
        & \vtfeed   & 94.1   &   0.0   \\
    \cmidrule{2-5}
      & \multirow{2}{*}{\slack{}~\cite{kreuk2019deceiving}} 
        & \sleipnir & 57.9   &   2.90  \\
      & 
        & \vtfeed   & 96.0   &   1.01  \\
    \midrule
    \multirow{9.2}{*}{Black-box} 
      & \multirow{2}{*}{\hdos{}~\cite{demetrio2019explaining}} 
        & \sleipnir &  0.0   &   0.0   \\
      & 
        & \vtfeed   &  0.0   &   0.0   \\
    \cmidrule{2-5}
      & \multirow{2}{*}{\hfield{}~\cite{nisi2021lost}} 
        & \sleipnir &  0.607 &   0.0   \\
      & 
        & \vtfeed   &  0.990 &   0.0   \\
    \cmidrule{2-5}
      & \multirow{2}{*}{\disp{}~\cite{lucas2021malware}} 
        & \sleipnir &  0.607 &   0.0   \\
      & 
        & \vtfeed   & 10.9   &   0.0   \\
    \cmidrule{2-5}
      & \multirow{2}{*}{\secinj{}~\cite{demetrio2021functionality}} 
        & \sleipnir & 99.2   &  99.6   \\
      & 
        & \vtfeed   & 76.2   & 100.0   \\
    \bottomrule
  \end{tabular}
  \end{center}
\end{table}

\section{Efficiency of RS-Del}
\label{app-sec:eval-comp-efficiency}

In this appendix, we discuss the training and computational efficiency of \rsdel. 
We provide comparisons with \rsabn~\cite{levine2020robustness}, which serves as a baseline in 
Appendix~\ref{app-sec:eval-cert-hamm-dist} for a more restricted Hamming distance threat model.

\paragraph{Computational efficiency}
Table~\ref{tbl:runtime-efficiency} reports wall clock times for training and prediction. 
For training, we measure the time taken to complete 1 epoch of stochastic gradient descent on the \sleipnir\ training 
set, where inputs are perturbed by the smoothing mechanism. 
For prediction, we measure the time taken for a single 1MB input file using $n_\mathrm{pred} = 1000$ Monte Carlo 
samples. 
We split the prediction time into two components: (1)~the time taken to generate perturbed inputs from the smoothing 
mechanism and (2)~the time taken to aggregate predictions for the perturbed inputs using the base MalConv model.
All times are recorded on a desktop PC fitted with an AMD Ryzen~7 5800X CPU and an NVIDIA RTX3090 GPU, using 
our PyTorch implementation of \rsdel\ and \rsabn. 
We execute training and prediction for the base model on the GPU, and the smoothing mechanism on the CPU. 
We use a single PyTorch process, noting that times may be improved by running the smoothing mechanism in parallel or 
on the GPU.

We now make some observations about the results. 
First, we note that training is an order of magnitude faster for \rsdel\ compared with \rsabn. 
We attribute this speed-up to the deletion smoothing mechanism of \rsdel, which drastically reduces the dimensionality 
of inputs, thereby reducing the time taken to perform forward and backward passes for the base model. 
On the contrary, the ablation smoothing mechanism of \rsabn\ does not alter the dimensionality of inputs, so 
it does not have a performance advantage in this respect.
Second, we observe that the total prediction time for \rsdel\ is approximately 150\% faster than for \rsabn. 
We expect this difference is also due to the effect of dimensionality reduction for the deletion smoothing mechanism.

\begin{table}
  \centering
  \caption{Comparison of runtime efficiency for two models: \rsdel\ (our method with byte-level deletion) and 
    \rsabn~\cite{levine2020robustness}. 
    The first column of wall times measures the time taken to train each model for one epoch on \sleipnir. 
    The second and third columns of wall times measure the time taken to make a prediction for a 1MB input file. 
    This is split into two components: the time taken to generate $n_\mathrm{pred} = 1000$ perturbed inputs from the 
    smoothing mechanism (second column) and the time taken to pass the perturbed inputs through the base model 
    (third column).
  }
  \label{tbl:runtime-efficiency}
  \begin{center}
    \small
    \begin{tabular}{
        ll
        d{4.0}
        d{2.2}
        d{1.3}
      } \toprule
               &                                & \multicolumn{3}{c}{Wall time (s)}                          \\
      \cmidrule(lr){3-5}
               &                                &                    & \multicolumn{2}{c}{Predict}           \\
      \cmidrule(lr){4-5}
      Model    & Parameters                     & \mc{Train 1~epoch} & \mc{Smoothing}   & \mc{Base model}    \\
      \midrule
      \rsdel   & $p_\del = 90\%$                &                354 & 10.42            & 0.070              \\
      \rsabn~\cite{levine2020robustness}   & $p_\ab = 90\%$                 &               1692 & 15.29            & 0.352              \\
      \midrule
      \rsdel   & $p_\del = 99\%$                &                329 &  8.79            & 0.043              \\
      \rsabn~\cite{levine2020robustness}   & $p_\ab = 99\%$                 &               1788 & 15.60            & 0.352              \\
      \bottomrule
    \end{tabular}
  \end{center}
\end{table}

\paragraph{Training efficiency}

\begin{figure}
  \centering
  \includegraphics[width=0.65\linewidth]{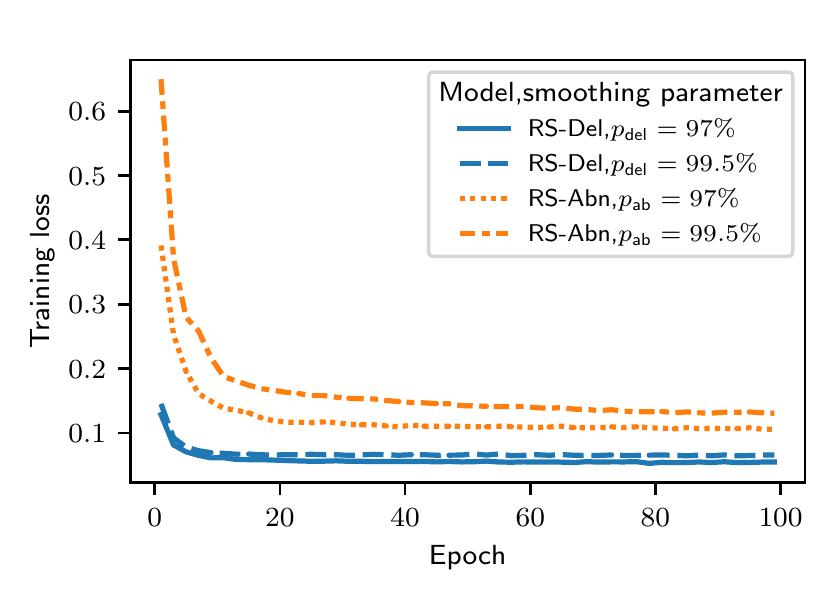}\\
  \caption{
    Training curves for \rsdel\ (our method with byte-level deletion) and \rsabn~\cite{levine2020robustness} for the 
    \sleipnir\ dataset. 
  }
  \label{fig:hist-sleipnir-hamming-cmp}
\end{figure}

Training curves for the base MalConv models used in \rsdel\ and \rsabn\ are provided in 
Figure~\ref{fig:hist-sleipnir-hamming-cmp} for the \sleipnir\ dataset. 
Due to convergence issues for \rsabn, we adapted training to incorporate gradient clipping when updating the 
embedding layer. 
This addresses imbalance in the gradients arising from the dominance of masked (ablated) values in the perturbed
inputs. 
However, even with this fix, we observe slower convergence to a higher loss value for \rsabn\ than for \rsdel. 
Combining the results of Table~\ref{tbl:runtime-efficiency} and Figure~\ref{fig:hist-sleipnir-hamming-cmp}, 
we conclude that \rsabn\ beats \rsabn\ in terms of training efficiency as it requires both fewer epochs to 
converge and takes less time per epoch.

\section{Adapting attacks for smoothed classifiers} \label{app-sec:adapt-wb}
In this appendix, we show how to adapt gradient-guided white-box attacks to account for randomized smoothing. 

We consider a generic family of white-box attacks that operate on neural network-based classifiers, where the first 
layer of the network is an embedding layer. 
Mathematically, we assume the classifier under attack $f$ can be decomposed as 
\begin{equation}
  f = f_\mathrm{embed} \circ f_\mathrm{soft} \circ f_\mathrm{pred}
  \label{eqn:attack-decomp}
\end{equation}
where 
\begin{itemize}
  \item $f_\mathrm{embed}$ is the embedding layer, which maps an input sequence $\vec{x} \in \Xspace$ of length 
  $n = \abs{\vec{x}}$ to an $n \times d$ array of $d$-dimensional embedding vectors;
  \item $f_\mathrm{soft}$ represents the subsequent layers in the network, which map an $n \times d$ embedding array 
  to a probability distribution over classes $\Yspace$; and
  \item $f_\mathrm{pred}$ is an optional final layer, which maps a probability distribution over classes to a 
  prediction (e.g., by taking the $\argmax$ or applying a threshold).
\end{itemize}
The attack may query any of the above components in isolation and compute gradients of $f_\mathrm{soft}$. 
For instance, the attacks we consider \cite{kreuk2019deceiving,demetrio2019explaining,lucas2021malware} optimize 
the input in the embedding space $\vec{e} \in \reals^{n \times d}$, by computing the gradient 
$\frac{\partial f_\mathrm{soft}(\vec{e})}{\partial \vec{e}}$. 

While the above setup covers attacks against MalConv, or classifiers with similar architectures, it is not directly 
compatible with smoothed classifiers. 
There are two incompatibilities.
First, a direct implementation of a smoothed classifier (as described in Section~\ref{sec:rs} and 
Figure~\ref{alg:estimation-algorithm}) does not decompose like \eqref{eqn:attack-decomp}.
And second, the aggregation of hard predictions from the base classifiers is non-differentiable. 
We explain how to address these incompatibilities below.

\paragraph{Leveraging commutativity}

Consider a smoothed classifier composed from a base classifier $h$ and smoothing mechanism $\phi$, and suppose the 
base classifier decomposes as in \eqref{eqn:attack-decomp}. 
Following Section~\ref{sec:rs} and Figure~\ref{alg:estimation-algorithm}, the smoothed classifier's confidence score 
for class $y$ can be expressed as
\begin{equation}
  p_y(\vec{x}) = \operatorname{smooth}_N(\vec{x}; \phi, 
  h_\mathrm{embed} \circ h_\mathrm{soft} \circ h_\mathrm{pred} \circ \ind{\square = y})
  \label{eqn:attack-smooth-original}
\end{equation}
where
\begin{equation*}
  \operatorname{smooth}_N(\vec{x}; \phi, f) = 
    \frac{1}{N} \sum_{i = 1}^{N} f(\vec{z}_i), \quad \text{with } \vec{z}_i \sim \phi(\vec{x})
\end{equation*}
is the (empirical) smoothing operation.
This expression does not immediately decompose like \eqref{eqn:attack-decomp}, because the embedding layer is applied 
to the perturbed input $\phi(\vec{x})$, not $\vec{x}$. 
Fortunately, we can manipulate the expression into the desired form by swapping the order of $\phi$ and 
$h_\mathrm{embed}$. 
To do so, we extend the definition of $\phi$ to operate on an embedding array so that 
$h_\mathrm{embed}(\phi(\vec{x})) = \phi(h_\mathrm{embed}(\vec{x}))$, i.e., $\phi$ and $h_\mathrm{embed}$ commute.
In particular, this can be done for randomized deletion (\rsdel) by applying the deletion edits to embedding vectors 
along the first dimension. 
Then \eqref{eqn:attack-smooth-original} can equivalently be expressed as
\begin{equation}
  p_y(\vec{x}) = \underbrace{h_\mathrm{embed}}_{f_\mathrm{embed}} \circ 
    \underbrace{\operatorname{smooth}_N(\phi, h_\mathrm{soft} \circ h_\mathrm{pred} \circ \ind{\square = y})}_{f_\mathrm{soft}}(\vec{x}).
  \label{eqn:attack-smooth-swap}
\end{equation}

\paragraph{Soft aggregation}
While \eqref{eqn:attack-smooth-swap} decomposes as required, the $f_\mathrm{soft}$ component is not differentiable. 
This is due to the presence of $h_\mathrm{pred}$, which is an $\argmax$ layer. 
To proceed, we replace the aggregation of predictions by the aggregation of softmax scores as proposed by  
\citeauthor{salman2019provably}~\cite{salman2019provably}. 
This yields a differentiable approximation of the smoothed classifier:
\begin{equation*}
  p_y(\vec{x}) \approx \underbrace{h_\mathrm{embed}}_{f_\mathrm{embed}} \circ 
    \underbrace{\operatorname{smooth}_N(\phi, h_\mathrm{soft} \circ \square_y)}_{f_\mathrm{soft}} (\vec{x}).
\end{equation*}
\citeauthor{salman2019provably} note that this approximation performs well, and is empirically more effective 
than an alternative approach proposed by \citeauthor{cohen2019certified}~\cite[Appendix~G.3]{cohen2019certified}.

\section{Review of randomized ablation}
\label{app-sec:rs-abn}

In this appendix, we review \emph{randomized ablation} \cite{levine2020robustness}, which serves as a baseline 
for the Hamming distance threat model in our experiments. 
It is based on \emph{randomized smoothing} (see Section~\ref{sec:rs}) like our method, however the smoothing mechanism 
and robustness certificate differ. 
In this review, we formulate randomized ablation for sequence classifiers; we refer readers to 
\citeteam{levine2020robustness} for a formulation for image classifiers. 

\subsection{Ablation smoothing mechanism}

Randomized ablation employs a smoothing mechanism that replaces a random subset of the input elements with a 
special null value $\NA$. 
\citeauthor{levine2020robustness} use an encoding for the null value tailored for images, that involves doubling 
the number of channels. 
This encoding is not suitable for discrete sequences, so we instead augment the sequence domain $\Omega$ with a 
special null value: $\Omega \to \Omega \cup \{\NA\}$.
The hyperparameter controlling the strength of ablation must also be adapted for our setting. 
\citeauthor{levine2020robustness} use a hyperparameter $k$ that corresponds to the number of elements \emph{retained} 
in the output. 
This is ineffective for inputs that vary in length, so we scale $k$ in proportion with the input length. 
Specifically, we introduce an alternative hyperparameter $p_\ab \in (0, 1)$ that represents the fraction of ablated 
elements and set $k(\abs{\vec{x}}) = \ceil{(1 - p_\ab) \abs{\vec{x}}}$. 

Mathematically, the ablation mechanism has the following distribution when applied to an input sequence $\vec{x}$:
\begin{equation*}
    \Pr[\phi(\vec{x}) = \vec{z}] = \sum_{\edit \in \Espace_{k(\abs{\vec{x}})}(\vec{x})} \frac{1}{\binom{\abs{\vec{x}}}{k(\abs{\vec{x}})}} 
        \ind{\ablate(\vec{x}, \edit) = \vec{z}},
\end{equation*}
where $\Espace_{k} = \{\edit \in \Espace(\vec{x}) : \abs{\edit} = k\}$ consists of all sets of element indices 
of size $k$ and 
\begin{equation*}
    z_i = \ablate(\vec{x}, \edit)_i = \begin{cases}
        x_i, & \text{if } i \in \edit \text{ ``retained''}, \\
        \NA, & \text{if } i \not\in \edit \text{ ``ablated''}. 
    \end{cases} 
\end{equation*} 

\begin{remark}
    Hyperparameter $p_\ab$ has a similar interpretation as $p_\del$ for randomized deletion, in that both 
    hyperparameters control the proportion of sequence elements hidden (by ablation or deletion) from the base 
    classifier.
\end{remark}

\subsection{Hamming distance robustness certificate}

Randomized ablation provides a Hamming distance ($\ell_0$) certificate that guarantees robustness under a bounded 
number of arbitrary substitutions. 
\citeauthor{levine2020robustness} provide two certificates: one that makes use of the confidence score for the 
predicted class, and another that makes use of the top two confidence scores.
We present the first certificate here, since it matches the certificate we consider in Section~\ref{sec:certificate} 
and it is the simplest choice for binary classifiers (the focus of our experiments).

To facilitate comparison with randomized deletion, we reuse notation and definitions from 
Section~\ref{sec:certificate}. 
Recall from \eqref{eqn:lb-cert} that $\tilde{\rho}(\vec{x}; \mu_y)$ is a lower bound on the 
smoothed classifier's confidence score $p_y(\pert{\vec{x}}; h)$ that holds for any input $\pert{\vec{x}}$ in the edit
distance ball of radius $\radius$ centered on $\vec{x}$ and any base classifier $h$ such that 
$p_y(\vec{x}; h) = \mu_y$. 
For randomized ablation, we can replace the edit distance ball with a Hamming distance ball and obtain the following 
result (see~\citep{levine2020robustness} for the derivation):
\begin{equation*}
    \tilde{\rho}(\vec{x}; \mu_y) = \mu_y - 1 + 
        \frac{\binom{\abs{\vec{x}} - \radius}{k(\abs{\vec{x}})}}{\binom{\abs{\vec{x}}}{k(\abs{\vec{x}})}}.
\end{equation*}
Recall from Proposition~\ref{prop:cert-suff}, that the smoothed classifier is certifiably robust if 
$\tilde{\rho}(\vec{x}; \mu_y) \geq \nu_y(\vec{\eta})$, where $\vec{\eta}$ denotes the smoothed classifier's tunable 
decision thresholds. 
Hence the certified radius $\radius^\star$ is the maximum value of $\radius \in \{0, 1, 2, \ldots\}$ that satisfies 
\begin{equation*}
    \mu_y - 1 + \frac{\binom{\abs{\vec{x}} - \radius}{k(\abs{\vec{x}})}}{\binom{\abs{\vec{x}}}{k(\abs{\vec{x}})}} - \nu_y(\vec{\eta}) \geq 0.
\end{equation*}
This maximization problem does not have an analytic solution, however it can be solved efficiently using binary 
search since the LHS of the above inequality is a non-increasing function of $\radius$. 
In practice, the exact confidence score $\mu_y$ can be replaced with a $1 - \alpha$ lower confidence bound $\umu_y$ 
to yield a probabilistic certificate that holds with probability $1 - \alpha$ (see procedure in 
Figure~\ref{alg:estimation-algorithm} and Corollary~\ref{cor:prob-cert}). 

Since randomized ablation and randomized deletion both admit Hamming distance certificates, it is interesting to 
compare them. 
The following result, first proved by \citeauthor{scholten2022randomized}~\cite[Proposition~5]{scholten2022randomized} 
for a related mechanism, shows that randomized deletion admits a tighter certificate.

\begin{proposition}
    Consider a randomized deletion classifier (\rsdel) with deletion probability $p_\del = p$ and a randomized 
    ablation classifier (\rsabn) with ablation fraction $p_\ab = p$. 
    Suppose both classifiers make the same prediction $y$ with the same confidence score $\mu_y$ for input 
    $\vec{x}$ and that we are interested in issuing a Hamming distance certificate of radius $\radius$. 
    Then the lower bound on the confidence score over the certified region is tighter for \rsdel than for \rsabn. 
\end{proposition}
\begin{proof}
    From Theorem~\ref{thm:lev-cert} we have
    \begin{align*}
        \tilde{\rho}_{\text{\rsdel}}(\vec{x}; \mu_y) 
        &= \mu_y - 1 + p^\radius \\
        &= \mu_y - 1 + \left(\frac{\abs{\vec{x}} - (1 - p) \abs{\vec{x}}}{\abs{\vec{x}}}\right)^\radius \\
        &\geq \mu_y - 1 + \left(\frac{\abs{\vec{x}} - \lceil (1 - p) \abs{\vec{x}} \rceil}{\abs{\vec{x}}}\right)^\radius \\
        &= \mu_y - 1 + \prod_{j = 1}^{\radius} \frac{\abs{\vec{x}} - k(\abs{\vec{x}})}{\abs{\vec{x}}} \\
        &\geq \mu_y - 1 + \prod_{j = 1}^{\radius} \frac{(\abs{\vec{x}} - k(\abs{\vec{x}}) - j + 1)}{(\abs{\vec{x}} - j + 1)} \\
        &= \mu_y - 1 + \frac{\binom{\abs{\vec{x}} - \radius}{k(\abs{\vec{x}})}}{\binom{\abs{\vec{x}}}{k(\abs{\vec{x}})}} \\
        &= \tilde{\rho}_\text{\rsabn}(\vec{x}; \mu_y)
    \end{align*}
\end{proof}

\begin{remark}
    \citeteam{jia2022almost} extend randomized ablation to top-$k$ prediction, while at the same time proposing 
    enhancements to the Hamming distance certificate. 
    These enhancements also apply to regular classification and involve: (1)~discretizing 
    lower\slash upper bounds on the confidence scores; and (2)~using a better statistical method to estimate 
    lower\slash upper bounds on the confidence scores. 
    However, both of these enhancements would have a negligible impact in our experiments, as we shall now explain. 
    The first enhancement tightens lower\slash upper bounds on the confidence scores by rounding up\slash down to the 
    nearest integer multiple of $q \coloneq 1/\binom{\abs{\vec{x}}}{k(\abs{\vec{x}})}$. 
    This improves the lower\slash upper bound by at most 
    \begin{equation*}
        q \leq \min \left\{
            \left(\frac{\abs{\vec{x}} - k}{\abs{\vec{x}}}\right)^{\abs{\vec{x}} - k}, 
            \left(\frac{k}{\abs{\vec{x}}}\right)^k 
        \right\} 
        \leq \frac{1}{\abs{\vec{x}}}
    \end{equation*}
    if the number of retained elements satisfies $1 \leq k \leq \abs{\vec{x}}$, as is the case in our experiments. 
    However, since we consider inputs of length $\abs{\vec{x}} \geq 10^3$, this means the improvement in the bound 
    due to discretization is at most $q \leq 10^{-3}$. 
    This improvement is comparable to the resolution of our estimator ($1 / n_\mathrm{bnd} \sim 10^{-4}$), and 
    consequently, discretization will not have a discernible impact. 
    The second enhancement involves simultaneously estimating lower\slash upper bounds using a statistical method 
    called SimuEM~\citep{jia2020certified}. 
    SimuEM has been demonstrated to improve tightness when there are multiple classes~\citep{jia2022almost}, however 
    it has no impact when there are two classes (as is the case in our experiments).
\end{remark}

\end{document}